\documentstyle[preprint,tighten,aps]{revtex}

\begin{document}

\draft 

\preprint{IFUP-TH 6/96}

\title{A strong-coupling analysis of two-dimensional
$\protect\bbox{{\rm O}(N)}$ $\protect\bbox{\sigma}$ models 
\protect\\
with $\protect\bbox{N\leq 2}$ on square, triangular and honeycomb
lattices}

\author{Massimo Campostrini, Andrea Pelissetto, Paolo Rossi, and
Ettore Vicari} 
\address{Dipartimento di Fisica dell'Universit\`a and I.N.F.N.,
I-56126 Pisa, Italy}

\maketitle

\begin{abstract}
The critical behavior of two-dimensional ${\rm O}(N)$ $\sigma$ models
with $N\leq 2$ on the square, triangular, and honeycomb lattices is
investigated by an analysis of the strong-coupling expansion of the
two-point fundamental Green's function $G(x)$, calculated up to 21st
order on the square lattice, 15th order on the triangular lattice, and
30th order on the honeycomb lattice.

For $N<2$ the critical behavior is of power-law type, and the
exponents $\gamma$ and $\nu$ extracted from our strong-coupling
analysis confirm exact results derived
assuming universality with solvable solid-on-solid models.

At $N=2$, i.e., for the 2-$d$ XY model, the results from all lattices
considered are consistent with the Kosterlitz-Thouless exponential
approach to criticality, characterized by an exponent
$\sigma=\case{1}{2}$, and with universality.  The value
$\sigma=\case{1}{2}$ is confirmed within an uncertainty of few per
cent.  The prediction $\eta=\case{1}{4}$ is also roughly verified.

For various values of $N\leq 2$, we determine some ratios of
amplitudes concerning the two-point function $G(x)$ in the critical
limit of the symmetric phase.  This analysis shows that the
low-momentum behavior of $G(x)$ in the critical region is essentially
Gaussian at all values of $N\leq 2$.  New exact results for the
long-distance behavior of $G(x)$ when $N=1$ (Ising model in the
strong-coupling phase) confirm this statement.
\end{abstract}



\section{Introduction}
\label{introduction}

The strong-coupling expansion is one of the most successful approaches
to the study of critical phenomena.  Many important results concerning
physical models at criticality have been obtained by deducing the
asymptotic critical behavior of physical quantities from their
strong-coupling series.

We have calculated the two-point Green's function
\begin{equation}
G(x)=\langle \,\vec{s}_x \cdot \vec{s}_0\,\rangle,
\label{gxf}
\end{equation}
of two-dimensional ${\rm O}(N)$ $\sigma$ models on the square,
triangular and honeycomb lattices, respectively up to 21st, 15th, and
30th order in the strong-coupling expansion. Such calculations were
performed within the nearest-neighbor lattice formulation, described
by the action
\begin{equation}
S= -N\beta \sum_{\rm links}  \vec{s}_{x_l}\cdot \vec{s}_{x_r},
\label{action}
\end{equation}
where $\vec{s}_x$ is a $N$-component vector, 
the sum runs over all the links, and $x_l,x_r$ indicate the sites
at the ends of each link. 
The comparison of results from strong-coupling series calculated
on different lattices  offers the possibility of important
tests of universality, which, if positive, strongly confirm the
reliability of the final results. 

A complete presentation of our strong-coupling computations for ${\rm
O}(N)$ $\sigma$ models in two and three dimensions will be presented
in a forthcoming paper.  A preliminary report on our calculations can
be found in Ref.~\cite{lattice95}.  On the square lattice our
strong-coupling series represent a considerable extension of the 14th
order calculations of Ref.~\cite{Luscher}, performed by means of a
linked cluster expansion, which have been rielaborated and analyzed in
Ref.~\cite{ButeraN3}.  We also mention recent works where the linked
cluster expansion technique has been further developed and
calculations of series up to 18th order~\cite{Reisz} and 19th
order~\cite{Butera2} for bulk quantities in $d=2,3,4$ have been
announced.
 
In this paper we focus on 2-$d$ ${\rm O}(N)$ $\sigma$ models with
$N\leq 2$.  The analysis of our strong-coupling series for models with
$N\geq 3$, i.e., those enjoying asymptotic freedom, is presented in
Ref.~\cite{ON-n3}.

Two-dimensional ${\rm O}(N)$ $\sigma$ models with $N<2$ should present
a standard power-law critical behavior, and should be described at
criticality by conformal field theories with central charge $c < 1$.
The most physically relevant models in this range of values of $N$ are
self-avoiding random walk models and Ising models, corresponding
respectively to $N=0$ and $N=1$.  At $N=-2$ the critical theory has
been proven to be Gaussian~\cite{Balian}, i.e., $\gamma=1$,
$\nu=\case{1}{2}$, and $\eta=0$.  Assuming universality with solvable
solid-on-solid models, exact formulas for the critical exponents in
the range $-2< N < 2$ have been
proposed~\cite{Nm2exp1,Nienhuis,Nm2exp2}, interpolating the critical
behaviors at $N=-2,0,1,2$.  The critical exponents of the magnetic
susceptibility $\gamma$ and correlation length $\nu$ are conjectured
to be
\begin{eqnarray}
\gamma&=&{3+a^2\over 4a(2-a)},\nonumber \\
\nu&=&{1\over 4-2a},
\label{nm2expf}
\end{eqnarray}
where the parameter $a$ is determined by the equation
\begin{equation}
N=-2\cos\left( {2\pi\over a}\right)
\end{equation}
with the constraint $1\leq a\leq 2$.  The exponent $\eta$ can be
obtained by the hyperscaling relation $\gamma=(2-\eta)\nu$.  

In the limit $N\rightarrow 2$ formulas (\ref{nm2expf}) yield
$\gamma\rightarrow\infty$ and $\nu\rightarrow\infty$, suggesting that
at $N=2$ the critical pattern should not follow a power-law behavior.
The XY spin model in two dimensions, i.e., the $N=2$ model, is
conjectured to experience a Kosterlitz-Thouless phase
transition~\cite{KT} of infinite order, characterized by a very weak
singularity in the free energy and an exponential divergence of the
correlation length at a finite $\beta$.  This model should describe
the critical properties of a number of two-dimensional systems, such
as thin films of superfluid helium.

According to the Kosterlitz-Thouless (K-T) scenario, the
correlation length is expected to behave like
\begin{equation}
\xi\sim \exp \left( {b\over \tau^{\sigma}} \right)
\label{xidiv}
\end{equation}
for $0<\tau\equiv 1-{\beta/\beta_c}\ll 1$.
The value of the exponent is  $\sigma=\case{1}{2}$ and $b$ is a 
non-universal positive constant. At the critical temperature, the
asymptotic behavior for $r\rightarrow\infty$ of the two-point
correlation function should be (cfr.\ e.g.\ Ref.~\cite{Itzykson})
\begin{equation}
G(r)_{\rm crit} \sim {(\ln r)^{2\theta}\over r^\eta}
\,\left[ 1 + O\left( {\ln\ln r\over \ln r}\right)\right],
\label{gx}
\end{equation}
with $\eta=\case{1}{4}$ and $\theta=\case{1}{16}$. Near criticality,
i.e., for $0 < \tau\ll 1$, the behavior of the magnetic susceptibility
can be deduced from Eq.~(\ref{gx}):
\begin{equation}
\chi\sim \int_0^\xi  dr \,G(r)_{\rm crit}\sim
\xi^{2-\eta} \left( \ln \xi \right)^{2\theta}
\left[ 1 + O\left( {\ln\ln \xi\over \ln \xi}\right)\right]\sim
 \xi^{2-\eta} \tau^{-2\sigma\theta}
\left[ 1 + O\left( \tau^{\sigma}\ln \tau\right)\right].
\label{chi}
\end{equation}
In addition, the 2-$d$ XY model is characterized by a line of critical
points, starting from $\beta=\beta_c$ and extending to $\beta=\infty$,
with $\eta$ going to zero as $1/\beta$ for $\beta\rightarrow\infty$.
At criticality the 2-$d$ XY model should give rise to a conformal
field theory with $c=1$.

Numerical studies based on Monte Carlo simulation techniques and
high-temperature expansions seem to support the K-T behavior, but a
direct accurate verification of all the K-T predictions is still
missing. As pointed out in Ref.~\cite{Hasenbush}, for $\beta<\beta_c$
the corrections to the asymptotic behavior (\ref{xidiv}) should become
really negligible only at very large correlation lengths, out of the
reach of standard Monte Carlo simulations on today's computers, which
allow $\xi\lesssim 100$ (cfr.\ Ref.~\cite{Gupta}, where simulations
for correlation lengths up to $\xi\simeq 70$ were performed on
lattices up to $512^2$).  Monte Carlo simulations supplemented with
finite-size scaling techniques allow to obtain data for larger $\xi$.
Ref.~\cite{Kim} shows data up to $\xi\simeq 850$, which, although
consistent with the K-T prediction, do not really exclude a standard
power-law behavior.\footnote{Actually the author of Ref.~\cite{Kim}
claims to favor a conventional power behavior to explain some
discrepancies in the determination of the critical exponent $\eta$.}

Finite-size scaling investigations at criticality are required to be
very precise in order to pinpoint the logarithm in the two-point
Green's function.  On the other hand, if this logarithmic correction
is neglected, the precise check of the prediction $\eta=\case{1}{4}$
at $\beta_c$ may be quite hard. The relevance of such logarithmic
corrections and some of the consequences of neglecting them have been
examined in Ref.~\cite{Irving}.
Numerical studies by Monte Carlo
renormalization-group and finite-size scaling
techniques~\cite{Petronzio,Gupta} seem to favor a lower value of
$\eta$, which might be caused by the neglected logarithm. The most
accurate verification of the K-T critical pattern has been shown in
Ref.~\cite{Hasenbush} by numerically matching the
renormalization-group trajectory of the dual of the XY model with that
of a the body-centered solid-on-solid model, which has been proven to
exhibit a K-T transition.  The advantage of this strategy is that such
a matching occurs much earlier than the onset of the asymptotic
regime, where numerical simulations can provide quite accurate
results.

The analysis of the strong-coupling series (cfr.\ e.g.\ 
Refs.~\cite{Butera,Buteratr}, where a few moments of the two-point
Green's function were calculated on the square and triangular lattices
respectively up to 20th and 14th order for the special value $N=2$)
substantially supports the K-T mechanism, but it does not provide
precise estimates for the exponents $\sigma$, $\eta$, and $\theta$,
probably for two reasons: (i) the asymptotic regime in the terms of
the series may be set at very large orders; (ii) the logarithmic
correction may cause systematic errors in most of the analysis
employed.

The computation of strong-coupling series on the honeycomb lattice
and the extension of series on the square and triangular lattices
motivate a new strong-coupling analysis of the 2-$d$ XY model.
We focus on the K-T mechanism, searching for evidences of this
phenomenon. 

As already shown in Refs.~\cite{ON-n3,ONgr}, the strong-coupling
analysis provides quite accurate continuum-limit estimates when
applied to dimensionless ratios of universal quantities, even in the
case of asymptotically free models, i.e., when the critical point is
$\beta_c=\infty$.  We define some dimensionless ratios of scaling
quantities (ratios of amplitudes) which characterize the low momentum
behavior of the two-point function $G(x)$, and estimate their values
in the critical regime by directly analyzing their strong-coupling
series.  This will allow us to check how close the low momentum
critical behavior of $G(x)$ is to Gaussian behavior.

The paper is organized as follows:

In Sec.~\ref{crbeh} we investigate the critical behavior of 2-$d$
${\rm O}(N)$ $\sigma$ models with $N\leq 2$ on the square, triangular
and honeycomb lattices, extracting the relevant critical parameters by
the analysis of the available strong-coupling series. For $N< 2$ we
compare the strong-coupling estimates of the critical exponents with
Eqs.~(\ref{nm2expf}). For $N=2$, i.e., the 2-$d$ XY model, we verify
the predictions of the K-T critical theory.  In particular
Sec.~\ref{analysis} presents the general features of our
strong-coupling analysis. Secs.~\ref{sub2}, \ref{sqan}, \ref{tran} and
\ref{hoan} contain detailed reports of the derivations of the various
results; they are rather technical and can be skipped by readers not
interested in the details of the analysis.  In Sec.~\ref{conc} the
principal results are summarized and some conclusions are drawn.

In Sec.~\ref{ratios} we evaluate, at criticality, the values of some
amplitude ratios concerning the low-momentum behavior of $G(x)$.  We
will present results for the most physically relevant models with
$N\leq 2$, i.e., those with $N=0,1,2$.  We also discuss their
implications on the low-momentum behavior of $G(x)$ in the critical
region of the symmetric phase.

In Sec.~\ref{Isingmodel} some new exact results concerning the
asymptotic large-distance behavior of $G(x)$ for the Ising models on
the square, triangular and honeycomb lattices are presented.

In Apps.~\ref{appa}, \ref{appb}, and \ref{appc} we present, for
$N=0,1,2$, the strong-coupling series of some relevant quantities used
in this study, respectively for the square, triangular, and honeycomb
lattice.

\section{Strong-coupling analysis of the critical behavior}
\label{crbeh}

\subsection{Analysis of the series}
\label{analysis}

From the Green's function $G(x)$ one can derive many interesting
quantities. Defining the moments of $G(x)$
\begin{equation}
m_{2j}= \sum_x (x^2)^j G(x),
\end{equation}
we computed on each lattice the magnetic susceptibility
$\chi$, and the second moment correlation length $\xi_G^2$,
\begin{eqnarray}
\chi&\equiv&m_0\;,\nonumber \\
\xi_G^2&\equiv &{m_2\over 4\chi}.
\label{chixi}
\end{eqnarray}

Models with $N<2$ should have a power-law critical behavior,
which may be appropriately investigated by analyzing the
strong-coupling series of $\chi$ and $\xi_G^2$, in order to extract
the critical exponents $\gamma$ and $\nu$.  For $N=2$, in order to
check the exponential approach to criticality predicted by the K-T
mechanism and extract the relevant exponent $\sigma$, as in
Ref.~\cite{Butera}, we consider the strong-coupling series of the
logarithm of $\chi$ and $\xi_G$.  More precisely, since
$\chi=1+O(\beta)$ and $\xi_G^2=\case{1}{4}c\beta + O(\beta^2)$, where
$c$ is the coordination number of the lattice ($c=4,6,3$ respectively
for the square, triangular and honeycomb lattice), we consider the
series
\begin{eqnarray}
&& l_\chi\equiv\beta^{-1}\ln \chi = 
   \sum_{i=0}^\infty c_i\beta^i,\nonumber \\
&& l_\xi\equiv\beta^{-1}\ln \left( {4\xi_G^2\over c\beta}\right) =
   \sum_{i=0}^\infty d_i \beta^i.
\label{lf}
\end{eqnarray}
According to Eqs.~(\ref{xidiv}) and (\ref{chi}) $l_\chi$ and $l_\xi$
should behave as  
\begin{equation}
l_\chi\sim l_ \xi \sim \tau^{-\sigma},
\label{asybeh}
\end{equation}
and are therefore suitable for a standard analysis by
Pad\'e or integral approximants.  
A vanishing 
exponent $\sigma$ would indicate a standard power-law critical
behavior.  Conversely a stable non-zero value of $\sigma$ would
exclude a power-law behavior.

Estimates of the critical exponents can be obtained by employing the
so-called critical point renormalization method (CPRM)~\cite{Baker}.
The idea is that, when
\begin{eqnarray}
A(x)&=&\sum_i a_i  x^i\sim (x_0-x)^{-\alpha}\nonumber \\
B(x)&=&\sum_i b_i x^i\sim (x_0-x)^{-\beta},
\label{ab}
\end{eqnarray}
we have
\begin{equation}
C(x)=\sum_i {b_i\over a_i}x^i\;\sim\;(1-x)^{-(1+\beta-\alpha)},
\label{cs}
\end{equation}
where now the position of the singularity is known. Therefore the
analysis of the series $C$ may provide an unbiased estimate for
the difference between the critical exponents of the two functions
$A$ and $B$. In particular this idea can be applied to the case
$B=A^2$, allowing one  to get a direct estimate of the critical
exponent of $A$, provided a  sufficiently large number of terms  is
known. The reliability of the determination of the critical
exponent by this method may be checked by comparing the results
for the critical point of an unbiased analysis with the exact
result $x_c=1$.

A general technique to extract physical information from a $n$th order
strong-coupling series $S(x)=\sum_{i=0}^n c_ix^i$ 
is constructing approximants $A(x)$ such that
\begin{equation}
A(x)-S(x)=O(x^{n+1}),
\end{equation}
and studying their singularities.  For a review on the resummation
techniques cfr.\ Ref.~\cite{Guttmann}.  $[l/m]$ Pad\'e approximants
(PA's) are ratios of two polynomials of degree $l$ and $m$
respectively, such that their Taylor expansion is equal to $S(x)$ up
to $O(x^{l+m})$.  PA's are expected to converge well to meromorphic
analytic functions. More flexibility is achieved by constructing PA's
of the logarithmic derivative of $S(x)$ (Dlog-PA analysis), and
therefore enlarging the class of functions which can be reproduced to
those having singularities of the form $(z-z_0)^\gamma$.  $[l/m]$
Dlog-PA's are obtained by integrating the $[l/m]$ PA's of the
logarithmic derivative of $S(x)$.  Then a $[l/m]$ PA uses $n=l+m$
terms of the series, while a $[l/m]$ Dlog-PA requires $n=l+m+1$ terms.

Other kind of approximants can be constructed as solutions of 
differential equations~\cite{IA}.
We consider integral approximants (IA's) obtained from a first
order linear differential equation
\begin{equation}
Q_m(x)f^\prime (x)+ P_l(x)f(x)+R_k(x)= O\left( x^{k+l+m+2}\right),
\label{intapprx}
\end{equation}
where $Q_m(x)$, $P_l(x)$ and $R_k(x)$ are polynomials of order $m$,
$l$, and $k$ respectively, and we fix $Q_m(0)=1$.  These approximants
are singular at the zeroes $x_0$ of $Q_m(x)$, and behave as
\begin{equation}
A(x)|x-x_0|^{-\gamma}+B(x),
\label{intapprx2}
\end{equation}
where $A(x)$ and $B(x)$ are regular in the neighborhood of $x_0$, and
\begin{equation}
\gamma = - {P_l(x_0)\over Q^\prime_m(x_0)}.
\end{equation}
When we analyze a $n$th order series, $m$, $l$ and $k$ must satisfy
the condition $k+l+m+2 \leq n\;$.  If the position of the singularity
$x_0$ is known, such an analysis can be easily modified forcing the
approximant to have a singularity at $x_0$ by substituting
$Q_m(x)\rightarrow (1-x/x_0)\bar{Q}_m(x)$, where $\bar{Q}_m(x)$ is
still a polynomial of order $m$ with $\bar{Q}_m(0)=1$.

Unlike Dlog-PA's, IA's are suited to take into account subdominant
terms in the vicinity of singularities, thus reducing possible
systematic errors in the resummation of the series.  On the other
hand, in order to get stable and therefore acceptable results, IA's
require in general more terms in the series to be resummed than PA's
or Dlog-PA's.

As a final estimate from each analysis we took the average of the
results from quasi-diagonal (non-defective) approximants (PA's or
IA's) using all available terms of the series. The errors we display
are just indicative, and should give an idea of the spread of the
results coming from the various approximants which can be constructed
from the series at hand.  They are the square root of the variance
around the estimate of the results coming also from quasi-diagonal
approximants constructed from shorter series by one and two terms.  In
the following we will specify the approximants considered in each
analysis.  This procedure does not always provide a reliable estimate
of the systematic error, which may be underestimated especially when
the structure of the function cannot be well reproduced by the class
of approximants used.  A more reliable estimate of the true
uncertainty should come from the comparison of results of different
analysis of the same series, and from the analyses of series of
different estimators of the same quantity, which in general are not
expected to have the same analytic structure.

\subsection{Critical behavior of models with $\protect\bbox{-2<N<2}$}
\label{sub2}

In order to determine the critical exponents $\gamma$ and $\nu$ of
2-$d$ ${\rm O}(N)$ $\sigma$ models with $N<2$, we analyze the
strong-coupling series of $\chi$ and $\xi_G^2$ on square (21st order),
triangular (15th order) and honeycomb (30th order) lattices.  For such
models, an analysis of the $14^{\rm th}$ order strong-coupling series
on the square lattice, calculated in Ref.~\cite{Luscher}, has been
done in Ref.~\cite{ButeraN3}.

For $N=0$ (the self-avoiding walk), longer series are available
\cite{GuttmannO0,GuttmannO0t}.  To compare with the literature,
observe that we have \cite{DeGennes}
\begin{equation}
G(x) = \sum_l \beta^l c_l(x),
\end{equation}
where $c_l(x)$ is the number of self-avoiding walks of length $l$
going from $0$ to $x$.  Therefore $\chi = \sum_l \beta^l c_l$, and 
\begin{equation}
\chi \xi_G^2 = \case{1}{4}\sum_l \beta^l c_l \left<R^2_e\right>_l,
\end{equation}
where $c_l = \sum_x c_l(x)$ is the total number of self-avoiding walk
of length $l$ starting from the origin, and
\begin{equation}
\left<R^2_e\right>_l = {1 \over c_l} \sum_x x^2 c_l(x)
\end{equation}
is called the ``mean end-to-end distance'' in self-avoiding walks
literature. 

Table~\ref{Nm2tab} shows the results of a Dlog-PA analysis, reporting,
for various values of $N$ and for each lattice, $\beta_c^{(\chi)}$ and
$\gamma$ as obtained from the strong-coupling series of $\chi$, and
$\beta_c^{(\xi)}$ and $\nu$ from that of $\xi^2_G$.  Differences
between $\beta_c^{(\chi)}$ and $\beta^{(\xi)}_c$ should give an idea
of the real uncertainty on $\beta_c$.  In the analysis of $\chi$ we
considered Dlog-PA's with $l+m\geq 18$ and $m\geq l\geq 8$ on the
square lattice, $l+m\geq 12$ and $m\geq l \geq 5$ on the triangular
lattice, $l+m\geq 27$ and $m\geq l \geq 12$ on the honeycomb lattice.
In the analysis of $\beta^{-1}\xi^2_G$ we considered Dlog-PA's with
$l+m\geq 17$ and $m\geq l\geq 8$ on the square lattice, $l+m\geq 11$
and $m\geq l \geq 5$ on the triangular lattice, $l+m\geq 26$ and
$m\geq l \geq 12$ on the honeycomb lattice.  We tried also IA's,
obtaining consistent results, which however only in few cases turned
out to be more precise than those of the Dlog-PA's, so we do not
report them.  For sake of completeness and also to give an idea of
the precision we can achieve with such an analysis, in
Table~\ref{Nm2tab} we report also results for $N=0,1$ as obtained from
our series, although exact results independent of the conjecture
(\ref{nm2expf}) exist for such values of $N$.  We warn that the errors
displayed in Table~\ref{Nm2tab} are related to the spread of the
results from the Dlog-PA's considered, according to the procedure
described in the Sec.~\ref{analysis}, and therefore they are not
always reliable estimates of the uncertainty.

In the range $-1\lesssim N \lesssim \case{3}{2}$ the formulas
(\ref{nm2expf}) for the exponents $\gamma$ and $\nu$ are well
reproduced, and universality is verified. Less precise determinations
are obtained when approaching the endpoints $N=\pm 2$, presumably due
to a rather slow convergence of the corresponding series to their
asymptotic regime.

We note that for models with $N\gtrsim 1$ on the honeycomb lattice the
physical critical point is not the singularity closest to the origin,
but there is a pair of closer singularities on the imaginary axis.
For instance, in the Ising model the physical singularity is placed at
$\beta_c={1\over 2}(2+\sqrt{3})=0.658478...$ and there is a pair of
singularities at $\bar{\beta} = \pm i \pi/6$~\cite{Shrock}.
Nevertheless Dlog-PA's of the magnetic susceptibility reproduce the
physical singularity very precisely: the [15/15] Dlog-PA gives
$\beta_c=0.658480$ and $\gamma=1.74993$, to be compared with the exact
result $\gamma=\case{7}{4}$. The unphysical singularities can be
mapped away from the origin by performing the change of variable
$\beta\rightarrow z={\rm tanh}\beta$, where $z$ is the character
coefficient of the fundamental representation.  With decreasing $N$,
$\beta_c$ decreases while the above-mentioned imaginary singularities
move away, so that at $N\lesssim 0$, the singularity closest to the
origin is on the real axis, i.e., it is the physical critical point.

For later comparison with the strong-coupling analysis of 2-$d$ XY
model, we have also analyzed the series of $l_\chi$, defined in
Eq.~(\ref{lf}), for the Ising model on the square lattice.  Since the
critical behavior is of power-law type, $l_\chi$ should have a
logarithmic singularity at $\beta_c$, and therefore an analysis like
IA should give $\sigma\simeq 0$~\cite{IA}. Indeed most of the IA's of
the 20th order series of $l_\chi$ give $|\sigma|\lesssim 0.02$. We
mention that a Dlog-PA analysis leads to misleading results in this
case, since in order to reproduce the logarithmic behavior it gives
rise to spurious singularities, which make the estimate of $\sigma$ at
$\beta_c$ unreliable.

\subsection{The 2-$\protect\bbox{d}$ XY model on the square lattice}
\label{sqan}

On the square lattice a strong-coupling analysis of the lowest moments
of $G(x)$ evaluated up to 20th order can be found in
Ref.~\cite{Butera}. Having achieved further extension by one term of
such series, we update here the situation on the square lattice.  We
note that the series obtained from our calculations (some of them are
reported in App.~\ref{appa}) present small discrepancies with those
reported in Ref.~\cite{Butera}: they are in agreement up to 16th
order, but slight different at higher order\footnote{The difference is
however small, at most $10^{-6}$, and it does not change the
conclusion of Ref.~\protect\cite{Butera}.}.  We are confident that our
series are exact, since they were generated for arbitrary $N$ and we
have checked their $N\rightarrow\infty$ limit against the exact
solution, and compared them with the existing series for $N=0,1$.

We analyzed the 20th order series of $l_\chi\equiv \beta^{-1}\ln \chi$
by both Dlog-PA's and IA's.  We found $\beta_c=0.560(2)$ and $\sigma=
0.53(4)$ from Dlog-PA's (with $l+m\geq 18$ and $m\geq l\geq 7$).  The
integral approximant analysis, whose details are given in
Table~\ref{lchisqIA}, leads to $\beta_c=0.558(2)$ and $\sigma=0.49(8)$
(considering IA's with $m+l+k\geq 19$ and $m\geq l,k \geq 5$).

From the strong-coupling series of $l_\chi$ and $l_\chi^2$, we have
constructed a new series $\lambda_\chi$ according to CPRM, and
analyzed it by standard methods: Dlog-PA's and IA's. We obtained
$\sigma=0.51(4)$ by Dlog-PA's (with $l+m\geq 18$ and $m\geq l\geq 8$)
biased by imposing the presence of a singularity at $x_c=1$, and
$\sigma=0.50(2)$ by biased IA's (with $m+l+k\geq 16$ and $m\geq l,k
\geq 5$).  Table~\ref{blchisqIA} shows some details of the IA analysis
of the series of $\lambda_\chi$.  From unbiased Dlog-PA's and IA's
$x_c$ is found to be equal to one within a few per mille, assuring us
on the reliability of the estimates of the exponent $\sigma$ by this
method.

The above unbiased analyses strongly support the K-T prediction
(\ref{chi}).  Although the estimate of $\sigma$ does not yet reach the
high level of precision which is usually found in the analysis of
strong-coupling series of considerable length, we can safely conclude
that the value $\sigma=\case{1}{2}$ is well verified with an
uncertainty of less than 10\% on the square lattice.

Unbiased approximants of $l_\xi$ give less 
stable but definitely consistent
results: we found $\sigma=0.59(6)$ from CPRM,
i.e., from an IA analysis biased at $x_c=1$
(with $m+l+k\geq 15$ and $m,l,k\geq 5$)
of the series $\lambda_\xi$ constructed
from $l_\xi$ and $l_\xi^2$ according to CPRM.
On the other hand  by applying CPRM to  the two series
$l_\chi$ and $l_\xi$ one may verify that
$\sigma_\chi=\sigma_\xi$. From an
IA analysis  biasing $x_c=1$ 
(with $m+l+k\geq 15$ and $m,l,k\geq 5$) we found
$\sigma_\chi-\sigma_\xi = 0.010(6)$, which represents
a good check of Eq.~(\ref{asybeh}).

Once $\sigma=\case{1}{2}$ was reasonably verified, we performed a set
of biased analysis fixing $\sigma=\case{1}{2}$ in order to determine
the critical point. A way to bias the value of the exponent at
$\sigma=\case{1}{2}$ is to analyze the series of the square of
$l_\chi$ and $l_\xi$ by PA's.  By doing so we obtained $\beta_c=
0.5579(3)$ from $l_\chi^2$, and $\beta_c=0.558(1)$ from $\l_\xi^2$ (we
used PA's with $l+m\geq 17$ and $m\geq l\geq 8$).  Still biasing
$\sigma=\case{1}{2}$, IA's yield $\beta_c=0.5583(2)$ form $l_\chi$,
and $\beta_c=0.559(1)$ from $l_\xi$ (here we determine, in a IA
analysis biasing the position of the singularity, the value of
$\beta_c$ which produces the exponent $\sigma=\case{1}{2}$.)

No complex singularities closer to the origin than $\beta_c$ 
are detected in the various strong-coupling analysis,
thus indicating that $\beta_c$ is also the convergence radius of
the strong-coupling expansion.

For $\beta<\beta_c$ we have compared $\chi$ and $\xi_G$ as obtained
from our strong-coupling analysis with numerical data, available in
the literature up to $\beta=1/1.96=0.5102...$ (corresponding to a
correlation length $\xi\simeq 70$) by using standard Monte Carlo
simulations~\cite{Gupta}, and up to $\beta=1/1.87=0.5347...$ by
employing also finite-size scaling techniques~\cite{Kim}.  Actually
most of the Monte Carlo data of $\xi$ reported in Ref.~\cite{Gupta}
concern $\xi_{\rm exp}$, i.e., the correlation length extracted from
the long distance exponential behavior of $G(x)$, but as we shall see
in the next section $\xi_G/\xi_{\rm exp} \simeq 0.999$ at criticality.
In order to get strong-coupling curves of $\chi$ and $\xi_G$ as
functions of $\beta$, we used approximants constructed biasing
$\sigma=\case{1}{2}$, i.e., PA's of $l_\chi^2$ and $l_\xi^2$, and IA's
of $l_\chi$ and $l_\xi$ biasing $\sigma=\case{1}{2}$.
Fig.~\ref{figchi} compares strong-coupling curves of $\ln \chi$ with
the available Monte Carlo data.  Table~\ref{mc} reports estimates of
$\xi_G$ by strong-coupling expansion and Monte Carlo simulations for
various values of $\beta$.  The agreement among the different
calculations is satisfactory.

In order to determine the exponent $\eta$ without biasing $\sigma$, we
considered various quantities which can be constructed using the
lowest moments of the two-point Green's function:
\begin{equation}
A_\eta\equiv {l_\chi\over l_\xi}=
{1\over 2}(1-\eta)+O(\tau^\sigma\ln\tau),
\label{Aeta}
\end{equation}
\begin{equation}
B_\eta\equiv (\beta l_\chi)^{-1}
\ln \left( 1+{m_2\over \chi^2}\right)
={\eta\over 2-\eta}+O(\tau^\sigma\ln\tau),
\label{Beta}
\end{equation}
\begin{equation}
C_\eta\equiv (\beta l_\xi)^{-1}
\ln \left( {m_2\over 4\beta\chi^2}\right)
={\eta\over 2}+O(\tau^\sigma\ln\tau).
\label{Ceta}
\end{equation}
An estimate of $\eta$ may then be obtained by resumming the
corresponding strong-coupling series by PA's and Dlog-PA's and
evaluating them at $\beta_c$. Since all of the above quantities are
equally good estimators of the exponent $\eta$, differences in the
results of their analysis should give an idea of the systematic error
in the procedure.

PA's and Dlog-PA's of $A_\eta$ (with $l+m\geq 16$ and $m\geq l\geq 7$)
lead to quite stable results: $\eta=0.228(2)$ (where the error
displayed, beside the spread of different approximants, takes into
account the uncertainty on $\beta_c$).  Similarly we found
$\eta=0.270(5)$ and $\eta=0.226(5)$ respectively from PA's and
Dlog-PA's of $B_\eta$ (with $l+m\geq 17$ and $m\geq l\geq 8$) and of
$\beta^{-1}C_\eta$ (with $l+m\geq 15$ and $m\geq l\geq 7$).  The
differences in such determinations indicate a systematic error of
about 10\%, and within 10\% all results are consistent with the K-T
prediction $\eta=\case{1}{4}$.  Furthermore, when analyzing the
energy-series of $A_\eta$ (by performing the change of variable
$\beta\rightarrow E$ and evaluating the corresponding approximants at
$E_c\simeq 0.722$, i.e., the energy value at $\beta_c\simeq
0.559$~\cite{Gupta}), we obtained again a rather stable result but
$\eta=0.207(5)$, confirming the presence of a systematic error of
about 10\%.

A source of systematic error in this analysis is the
$O(\tau^\sigma\ln\tau)$ correction expected in Eqs.~(\ref{Aeta}),
(\ref{Beta}) and (\ref{Ceta}) which cannot be reproduced by PA's and
Dlog-PA's. In particular Eq.~(\ref{Aeta}) implies a behavior
\begin{equation}
{\rm Dlog} A_\eta\sim\tau^{\sigma-1}
\label{err}
\end{equation}
in the proximity of to $\beta_c$. In the Dlog-PA's the above
singularities should be mimicked by a pole shifted at a $\beta$ larger
than $\beta_c$. Indeed in the analysis of the series of $A_\eta$ we
have found a singularity typically at $\beta\simeq 1.1\div
1.2\beta_c$. This fact will eventually affect the determination of
$A_\eta$ close to $\beta_c$ by a systematic error. However, since the
singularity is integrable, the error must be finite, and the analysis
shows that such errors are actually reasonably small. The behavior
(\ref{Aeta}), neglecting the logarithm, could be reproduced by IA's ,
but we did not obtain sufficiently stable and therefore acceptable
results by them. As a further check we also performed a biased Dlog-PA
analysis of the series of the quantity $A_\eta - {1\over
2}(1-\eta)\sim \tau^{\sigma}$ (neglecting logarithmic corrections). By
fixing $\eta=\case{1}{4}$ and $\beta_c=0.559$ we found an exponent
$\sigma\simeq 0.6$, which is satisfactorily close to the expected
value $\sigma=\case{1}{2}$.

The exponent $\theta$ defined in Eqs.~(\ref{gx}) and (\ref{chi}) may
be extracted by the analysis of the series of the ratios ${\chi/
\xi_G^{2-\eta}}$ and ${m_4/\xi_G^{6-\eta}}$, indeed
\begin{equation}
{\chi\over \xi_G^{2-\eta}}\sim {m_4\over \xi_G^{6-\eta}}
\sim\tau^{-2\sigma \theta}\left[ 1 + O(\tau^{\sigma}\ln \tau)\right].
\label{theta}
\end{equation}
Fixing $\eta=\case{1}{4}$, we performed biased analyses of the 20th
order strong coupling series of the above ratios imposing
$\beta_c=0.559$.  Dlog-PA's (with $l+m\geq 17$ and $m,l\geq 8$)
provide the following estimates for $\theta$ (obtained taking
$\sigma=\case{1}{2}$): $\theta = -0.042(5)$ from ${\chi/
\xi_G^{2-\eta}}$, and $\theta = -0.05(2)$ from ${m_4/
\xi_G^{6-\eta}}$, where errors take into account, beside the spread of
the Dlog-PA results, also the uncertainty on $\beta_c$.  These
numbers, although they confirm the fact that $|\theta|$ is small, are
rather different from the K-T prediction $\theta=\case{1}{16}$.  As
already mentioned above, a source of systematic error for a Dlog-PA
analysis is the correction $O(\tau^{1/2}\ln\tau)$ to the leading
$\tau^{-\theta}$ behavior in formula (\ref{theta}).  Since $\theta$ is
very small, this could cause a relevant departure from its true value.
The analysis by IA's biased at $\beta_c\simeq 0.559$ of the strong
coupling series of the ratio ${\chi/ \xi_G^{2-\eta}}$ does not provide
sufficiently stable results for $\theta$.

\subsection{The 2-$\protect\bbox{d}$ XY model on the triangular
lattice}
\label{tran}

On the triangular lattice strong-coupling series of some lowest
moments of $G(x)$ have been calculated in Ref.~\cite{Buteratr} up to
14th order. We calculated $G(x)$ up to 15th order, thus extending by
one order earlier calculations.  We must again mention the existence
of discrepancies between our calculations (cfr.\ App.~\ref{appb}) and
those of Ref.~\cite{Buteratr} in the 14th order terms (again of the
order of $10^{-6}$).

We performed on the triangular lattice the kind of analysis presented
in the previous Subsection for the square lattice.  We analyzed the
14th order series of $l_\chi\equiv\beta^{-1}\ln \chi$ by both
Dlog-PA's and IA's.  We found $\beta_c=0.3413(3)$, $\sigma= 0.52(2)$
from Dlog-PA's (with $l+m\geq 11$ and $m\geq l\geq 5$), and
$\beta_c=0.33986(4)$, $\sigma= 0.473(3)$ from IA's with $m+l+k\geq 12$
and $m\geq l,k\geq 3$ (the apparent stability of these IA
determinations should not be taken seriously, we remind again that
what it is really important for estimating the uncertainty is the
comparison of results from different analysis).

The CPRM applied to the strong-coupling series of $l_\chi$ and
$l_\chi^2$ gives $\sigma=0.53(1)$ by biased Dlog-PA's having a
singularity fixed at $x_c=1$ (with $l+m\geq 12$ and $m\geq l\geq 5$)
and $\sigma=0.50(2)$ by biased IA's (with $m+l+k\geq 11$ and $m\geq
l,k\geq 3$).  Table~\ref{blchitrIA} shows some details of the IA
analysis.  From unbiased Dlog-PA's and IA's, $x_c$ is found to be
equal to one within a few per mille, assuring us on the reliability of
the estimates of the exponent $\sigma$ by this method.

The critical point renormalization method gives good results also when
applied to the series of $l_\xi$, leading to $\sigma=0.52(4)$.  By
applying CPRM to the two series $l_\chi$ and $l_\xi$ one finds
$\sigma_\chi-\sigma_\xi=0.001(2)$ from an IA analysis biased at
$x_c=1$ (with $m+l+k\geq 12$ and $m\geq l,k\geq 3$).

We can conclude that the K-T prediction $\sigma=\case{1}{2}$
is strongly supported by the above results.

We also performed a set of biased analysis fixing $\sigma=\case{1}{2}$
in order to determine the critical point.  By analyzing $l_\chi^2$ and
$l_\xi^2$ by PA's, we obtained $\beta_c= 0.3400(2)$ and
$\beta_c=0.3393(1)$ respectively (from PA's with $l+m\geq 12$ and
$m\geq l\geq 5$).  IA's of $l_\xi$ biased at $\sigma=\case{1}{2}$
yield $\beta_c=0.3410(5)$, while when applied to $l_\chi$ such
analysis does not lead to relevant results, since it gives rise to
spurious singularities in the real axis.  A $\sigma=\case{1}{2}$
biased estimate of the critical point is then $\beta_c=0.340(1)$.

We obtained estimates of $\eta$ by resumming the series of $A_\eta$
and $B_\eta$ (cfr.\ Eqs.~(\ref{Aeta}) and (\ref{Beta})) by PA's and
Dlog-PA's and evaluating them at $\beta_c\simeq 0.340$. PA's and
Dlog-PA's of $A_\eta$ (with $l+m\geq 10$ and $m\geq l\geq 5$) and of
$B_\eta$ (with $l+m\geq 11$ and $m\geq l\geq 5$) lead again to quite
stable but slightly discrepant results: respectively $\eta=0.221(2)$
and $\eta=0.270(4)$.  The causes of possible systematic errors in the
determination of $\eta$ are the same as for the square lattice, and we
refer to Subs.~\ref{sqan} for a discussion.

The analysis of the 14th order strong coupling series of
$\chi/\xi_G^{2-\eta}\sim\tau^{-\theta}$ biased by $\beta_c\simeq
0.340$ (using Dlog-PA's with $l+m\geq 11$ and $l,m\geq 5$) yields the
estimate $\theta=-0.045(3)$, which is consistent with the square
lattice result, but not with the K-T prediction.  IA's does not
provide sufficiently stable results also in this case.

\subsection{The 2-$\protect\bbox{d}$ XY model on the honeycomb
lattice}
\label{hoan}

On the honeycomb lattice we calculated series longer than on the
square lattice, up to 30th order.  Here the possibility
of reaching larger orders is related to the smaller coordination
number. However longer series do not necessarily mean that more
precise results can be obtained from their analysis. This
possibility is related to the approach to the asymptotic regime of
the series, which is expected to be set later on lattice 
with smaller coordination number. Actually,  as we shall see, the
30th order series on the honeycomb lattice provide
results consistent with the K-T theory and universality, but  less
precise that those obtained from the series on the square 
and triangular lattices.

Unbiased analyses of the series for $l_\chi$ lead to:
$\beta_c=0.884(1)$ and $\sigma= 0.55(1)$ from Dlog-PA's (with $l+m\geq
27$ and $m\geq l\geq 12$), and $\beta_c=0.877(6)$, $\sigma= 0.4(2)$
from IA's (with $m+l+k\geq 27$ and $m\geq l,k\geq 8$).  The stability
of Dlog-PA's is suspect in this case, indeed we found that, just by
adding to the series a simple constant of the order of unity, the
change in the estimate of $\sigma$ turns out to be much larger than
the error evaluated from the spread of the estimates of different
approximants of the same series. Unlike PA's and IA's, the critical
parameters provided by Dlog-PA's do not remain strictly invariant by
adding a simple constant; this is only an asymptotic property of the
Dlog-PA analysis.  An analysis based on CPRM method fails to give
stable results in this cases, probably because the available series
are not sufficiently long to have their asymptotic regime set.

More stable results are obtained when biasing the exponent at
$\sigma=\case{1}{2}$. A check of consistency would be requiring that
the critical points as extracted from $l_\chi$ and $l_\xi$ are the
same. We obtained $\beta_c=0.879(1)$ from PA's (with $l+m\geq 26$ and
$m\geq l\geq 13$) of $l_\chi^2$, and $\beta_c=0.878(2)$ from PA's of
$l_\xi^2$, which is satisfactory.

It is worth noticing that, unlike what happens on the square and
triangular lattices, on the honeycomb lattice the real singularity
corresponding to the critical behavior of the theory is not the
singularity closest to the origin in the complex $\beta$-plane. A pair
of imaginary singularities at $\beta\simeq \pm i0.482$ is detected in
the analysis of the strong-coupling series of $\chi$.

Taking as an estimate of the critical point $\beta_c\simeq 0.880$, we
evaluated $\eta$ from the series of $A_\eta$ and $B_\eta$ defined in
Eqs.~(\ref{Aeta}) and (\ref{Beta}). Again the value obtained from
$A_\eta$ is about 10\% lower than $\case{1}{4}$: $\eta=0.231(3)$ (from
PA's and Dlog-PA's with $l+m\geq 26$ and $m\geq l\geq 11$), and that
from $B_\eta$ is about 10\% higher: $\eta=0.28(1)$ (from PA's and
Dlog-PA's with $l+m\geq 27$ and $m\geq l\geq 12$).  The behavior of
the estimates from $A_\eta$ and $B_\eta$ observed in the various
lattice seems to indicate that the source of systematic error is in a
sense universal, i.e., it essentially depends on the quantity
considered and approximately independent of the lattice.

We again estimated the exponent $\theta$ from the 29th order strong
coupling series of the $\chi/\xi_G^{2-\eta}$.  Dlog-PA's (with
$l+m\geq 26$ and $l,m\geq 13$) biased at $\beta_c\simeq 0.880$ give
$\theta=-0.042(6)$, which is consistent with the estimates from the
other lattices.

\subsection{Conclusions}
\label{conc}

We have studied the critical properties of 2-$d$ ${\rm O}(N)$ $\sigma$
models with $N\leq 2$ on the square, triangular and honeycomb
lattices, by analyzing the strong-coupling expansion of the lowest
moments of the two-point fundamental Green's function.

The analysis of the strong-coupling series of $\chi$ and $\xi_G^2$ on
the square, triangular and honeycomb lattices has substantially
confirmed that models with $N<2$ present a standard power-law critical
behavior with critical exponents given by Eqs.~(\ref{nm2expf}).  We
obtained rather precise determinations of the critical exponents in
the region $-1\lesssim N\lesssim \case{3}{2}$ (cfr.\ 
Table~\ref{Nm2tab}), where formulas (\ref{nm2expf}) are verified
within 1\%. The strong-coupling analysis becomes less precise
approaching the endpoints $N=\pm 2$, presumably due to a rather slow
convergence of the corresponding series to their asymptotic regime.
Universality among models on the square, triangular and honeycomb
lattices has been verified.

The determinations of $\beta_c$ and $\sigma$ for the 2-$d$ XY model on
the three different lattices are summarized in Table~\ref{summary}.
These results are consistent with the K-T exponential approach to
criticality and with universality.  The best estimates of $\sigma$
come from the analysis of the series of the magnetic susceptibility,
leading to a confirm of the value $\sigma=\case{1}{2}$ within an
uncertainty of few per cent.  The analysis of the series of the
correlation length $\xi_G^2$ yields consistent results.  The critical
point renormalization method~\cite{Baker} provides the most precise
unbiased estimates of $\sigma$ on the square and triangular lattices.
These results rule out the possibility of a standard power-law
critical behavior.

On the square lattice most estimates of $\beta_c$ yielded by our
analyses lie in the range $0.558\leq \beta_c \leq 0.560$, although the
lowest value $\beta_c=0.558$ seems to be favored.  This value is
consistent with the results of an exponential fit to data of $\xi$ up
to $\xi\simeq 850$~\cite{Kim}, which yielded $\beta_c=0.5593(13)$ and
$\sigma=0.46(3)$, and with a biased exponential fit fixing
$\sigma=\case{1}{2}$ to data up to $\xi\simeq 70$ produced by a
standard Monte Carlo simulations~\cite{Gupta}, which gave
$\beta_c=0.559(3)$.  But it is slightly smaller than the quite precise
Monte Carlo renormalization group determination of
Ref.~\cite{Hasenbush}: $\beta_c=0.55985(25)$.  The comparison of
(suitable) resummations of the strong-coupling series of $\chi$ and
$\xi_G^2$ with Monte Carlo data (available up to $\xi\simeq 850$)
turns out to be quite satisfactory (cfr.\ Table~\ref{mc}), giving
further support to our conclusions.

The prediction $\eta=\case{1}{4}$ is also substantially verified.  By
using different estimators of $\eta$ we control the systematic error
of our analysis, which turns out to be about 10\%, and within about
10\% our estimates of $\eta$ are always consistent with the value
$\eta=\case{1}{4}$.

Substantial discrepancies from the Kosterlitz-Thouless theory are
found in the estimates of $\theta$ we obtained on all lattices
considered and using also different estimators.  Our strong coupling
analysis based on Dlog-PA's leads, similarly to the K-T prediction, to
a small absolute value of $\theta$, but it would favor the value
$\theta\simeq -0.04$, against the K-T value $\theta=\case{1}{16}$.
Our strong coupling estimate seems to pass the universality check by
changing lattice and estimator.  On the other hand, we suggest some
caution in considering our strong coupling estimate of $\theta$.
Given the smallness of its value, we cannot exclude that the observed
discrepancy is due to systematic errors caused by the fact that
Dlog-PA's cannot reproduce the correction $O(\tau^{1/2}\ln\tau)$ to
the leading $\tau^{-\theta}$ behavior in formula (\ref{theta}).  Since
this correction is expected to be present in all quantities we
considered to estimate $\theta$ (even when defined on different
lattices), if its coefficients in the various cases are quantitatively
similar the error might be about the same, and explain the apparent
universality of our results.  Moreover, the more general analysis
based on IA's does not provide sufficiently stable results when
applied to estimate $\theta$, likely because the available series are
not sufficiently long to this purpose.  We mention that in
Ref.~\cite{Irving} an analysis based on Monte Carlo simulations led to
the estimate $\theta\simeq 0.02$, which is not consistent with both
the K-T prediction and our strong coupling estimate.

\section{Low-momentum behavior of $\protect\bbox{G(\lowercase{x})}$ in
the critical region} 
\label{ratios}

In this section we study the low-momentum behavior of the two-point
fundamental Green's function in the critical limit of the symmetric
phase. To this purpose, we consider the dimensionless
renormalization-group invariant function
\begin{equation}
L(p;\beta)\equiv{\widetilde{G}(0;\beta)\over\widetilde{G}(p;\beta)}.
\label{ldef}
\end{equation}
In the critical region of the symmetric phase $L(p,\beta)$ is a
function of the ratio $y\equiv p^2/M_G^2$ only, where $M_G\equiv
1/\xi_G$ and $\xi_G\equiv m_2/4\chi$ is the second moment correlation
length, already introduced in the previous section.  $L(y)$ can be
expanded in powers of $y$ around $y=0$:
\begin{eqnarray}
L(y)&=&1 + y + l(y)\nonumber \\
l(y)&=&\sum_{i=2}^\infty c_i y^i.
\label{lexp}
\end{eqnarray}
$l(y)$ parameterizes the difference from a generalized Gaussian
propagator.  The coefficients $c_i$ of the low-momentum expansion of
$l(y)$ can be related to appropriate dimensionless
renormalization-group invariant ratios of moments $m_{2j}=\sum_x
(x^2)^j G(x)$. Let us introduce the quantities
\begin{equation}
v_{2j}={1\over 2^{2j} (j!)^2 }M_G^{2j}{m_{2j} \over m_0} 
\label{vj}
\end{equation}
whose continuum limit is
\begin{equation}
v_{2j}^*={(-1)^j \over j!} {d\over dy^j} L(y)^{-1}|_{y=0}.
\label{vjstar}
\end{equation}  
$v_{2j}^*=1$ for a Gaussian critical propagator.
One can easily write the coefficients $c_i$ in terms of $v_{2j}^*$:
\begin{eqnarray}
c_2&=& 1-v_4^*,\nonumber \\
c_3&=& 1-2v_4^*+v_6^*\nonumber \\
c_4&=& 1+v_4^*(v_4^*-3)+2v_6^*-v_8^*,
\label{cvj}
\end{eqnarray}
etc. The strong-coupling expansion of $G(x)$ allows to calculate
strong-coupling series of $v_{2j}$. Estimates of the coefficients
$c_i$ can then be obtained, as we shall see, from the analysis of the
combinations of $v_{2j}$ corresponding to the r.h.s.\ of (\ref{cvj}).

Another quantity which characterizes the low-momentum behavior of
$L(y)$ is the ratio $s=M^2/M_G^2$ where $M$ is the mass-gap of the
theory, i.e., the mass determining the long distance exponential
behavior of $G(x)$.  The values $s^*$ of $s$ in the critical limit is
related to the zero $y_0$ of $L(y)$ closest to the origin: indeed
$y_0=-s^*$.  $s^*$ is in general different from one; it is one in
Gaussian-like models (i.e., when $l(y)=0$), such as the large-$N$
limit of ${\rm O}(N)$ $\sigma$ models, while no exact results are
known at finite $N$.

In the absence of a strict rotation invariance, one may actually
define different estimators of the mass-gap having the same continuum
limit.  On the square lattice one may consider $\mu$ obtained by the
long distance behavior of the side wall-wall correlation constructed
with $G(x)$, or equivalently the solution of the equation
$\widetilde{G}^{-1}(p_1=i\mu,p_2=0)=0$. In view of a strong-coupling
analysis, it is convenient to use another estimator of the mass-gap
derived from $\mu$:
\begin{equation}
M_{\rm s}^2= 2\left( {\rm cosh} \mu - 1\right),
\label{MM}
\end{equation}
which has an ordinary strong-coupling expansion
\begin{equation}
M_{\rm s}^2={1\over\beta}\left( 1+\sum_{i=1}^\infty a_i\beta^i\right)
\label{MM2}
\end{equation}
($\mu$ has a singular strong-coupling expansion, starting with
$-\ln\beta$).  One can easily check that $M_{\rm s}/\mu\rightarrow 1$
in the critical limit.  Similar quantities $M^2_{\rm t}$ and $M^2_{\rm
h}$ can be defined respectively on the triangular and honeycomb
lattices, as shown in Apps.~\ref{appb} and \ref{appc}.  One may then
consider the dimensionless ratios $M_{\rm s}^2/M_G^2$, $M_{\rm
t}^2/M_G^2$ and $M_{\rm h}^2/M_G^2$ respectively on the square,
triangular and honeycomb lattices, and evaluate their fixed point
limit $s^*$, which by universality must be the same for all of them.
From the available strong-coupling series of $M^2_{\rm s}$ and $M^2_G$
on the square lattice, $M^2_{\rm t}$ and $M^2_G$ on the triangular
lattice, $M^2_{\rm h}$ and $M^2_G$ on the honeycomb lattice, which are
reported, for $N=0,1,2$, in Apps.~\ref{appa},~\ref{appb},~\ref{appc}
respectively, we computed the ratio $M^2_{\rm s}/M^2_G$ up to 16th
order, $M^2_{\rm t}/M^2_G$ up to 11th order, $M^2_{\rm h}/M^2_G$ up to
25th order.
For the Ising models, using the known exact results for
$M_{\rm s}^2$, $M_{\rm t}^2$, and $M_{\rm h}^2$ (see next section),
we obtained longer series, i.e. 
$M^2_{\rm s}/M^2_G$ up to 20th
order, $M^2_{\rm t}/M^2_G$ up to 14th order, $M^2_{\rm h}/M^2_G$ up to
29th order.

In order to determine $s^*$, and the coefficients $c_i$ of the
low-momentum expansion of $L(y)$, we analyzed the strong-coupling
series of the ratios $M_{\rm s}^2/M_G^2$, $M_{\rm t}^2/M_G^2$ and
$M_{\rm h}^2/M_G^2$, and of the combinations of $v_{2j}$ given in
Eq.~(\ref{cvj}).  Beside the ordinary series in $\beta$, we also
considered and analyzed the corresponding series in the energy.  The
change of variable from $\beta$ to the energy $E$ is easily performed
by inverting the strong-coupling series of the energy
$E=\beta+O(\beta^3)$ and substituting into the original series in
powers of $\beta$.  We constructed PA's and Dlog-PA's (and sometimes
as further check also IA's) of both the series in $\beta$ and in $E$.
While PA's provide directly the quantity at hand, in a Dlog-PA
analysis one gets corresponding approximants by reconstructing the
original quantity from the PA of its logarithmic derivative.
Estimates at criticality are then obtained by evaluating the
approximants of the $\beta$-series at $\beta_c$, and those of the
$E$-series at $E_c$, i.e., the value of the energy at $\beta_c$.  In
the cases in which $E_c$ is not known from independent studies, its
estimate may be derived from the first real positive singularity
detected in the analysis of the strong-coupling series of $\chi$, or
$l_\chi$ for $N=2$, expressed in powers of $E$.

In our analysis we considered quasi-diagonal $[l/m]$ PA's and
Dlog-PA's of the available series; more precisely, for a $n$th order
series we considered those with
\begin{equation}
l,m\geq  {n\over 2} - 2,\;\;\;\;\;\;\;\;
l+m\geq n-2.
\label{paap}
\end{equation}
As a final estimate from each analysis we take the average of the
results from the quasi-diagonal PA's and Dlog-PA's using all available
terms of the series. The errors we display are the square root of the
variance around the estimate of the results from all non-defective
PA's indicated by Eq.(\ref{paap}). 

By analyzing the above-mentioned series at $N=0,1,2$ we obtained
estimates of $s^*$, and of some of the coefficients $c_i$.  The
results for $N=2$, i.e., for the XY model, are reported in
Table~\ref{tabsuN2}, those for $N=1,0$ in Table~\ref{tabsuN10}.
Universality among the square, triangular and honeycomb lattices is in
all cases well verified and gives further support to our final
estimates.

For the XY model, the analysis of the $E$-series provides the most 
precise results on all lattices considered, leading to the estimates:
\begin{eqnarray}
s^* &=& 0.9985(5),\nonumber \\
c_2 &=& -1.5(5)\times 10^{-3},\nonumber \\ 
c_3 &=& 2(2)\times 10^{-5}.
\label{resn2}
\end{eqnarray}
The errors displayed are a 
rough estimate of the uncertainty.

For the Ising model, the two-point function in the scaling region is
known analytically~\cite{Wu-McCoy}.  We obtained a benchmark for our
strong-coupling computation by computing numerically the two-point
function, following Ref.~\cite{Wu-McCoy}, and performing a numerical
integration of the results:
\begin{eqnarray}
s^* &\cong& 0.99919633,\nonumber \\
c_2 &\cong&-0.7936796\times 10^{-3},\nonumber \\
c_3 &\cong& 1.095991\times 10^{-5},\nonumber \\
c_4 &\cong&-3.12747\times 10^{-7},\nonumber \\
\label{exresn1}
\end{eqnarray}
The analysis of the available strong coupling series on the
square, triangular and honeycomb lattices
lead to the final estimates:
\begin{eqnarray}
s^* &=& 0.99908(3),\nonumber \\
c_2 &=&-0.94(4)\times 10^{-3},\nonumber \\
c_3 &=& 1.1(3)\times 10^{-5}.
\label{resn1}
\end{eqnarray}
The agreement with the exact results (\ref{exresn1}) is satisfactory.
But the comparison shows also that the errors on the strong coupling
estimates of $s^*$ and $c_2$, essentially calculated using the
variance of results from different PA's, are underestimated.  We
mention an earlier attempt to estimate $s^*$ for the Ising model by
using shorter strong-coupling series on the square and triangular
lattices~\cite{Fisher}.

For the self-avoiding random walk model we find:
\begin{eqnarray}
s^* &=& 1.0000(2),\nonumber \\
c_2 &=& 0.13(6)\times 10^{-3},\nonumber \\
|c_3| &\lesssim& 2\times 10^{-5}.
\label{resn0}
\end{eqnarray}

The analysis of the coefficients $c_i$ with $i > 3$ becomes
less and less precise with increasing $i$, 
but it is consistent with very small values.
For instance, we found in all cases  $|c_4|\lesssim |c_3|$.

So, for all $N$ considered, our strong-coupling analysis leads to the
following pattern of the coefficients $c_i$
\begin{equation}
c_i\ll c_2\ll 1 \;\;\;\;\;\;\;\;\;{\rm for}\;\;\;\; i\geq 3.
\label{crel}
\end{equation}  
This was also observed in models with $N\geq 3$ by a study based on
large-$N$ and strong-coupling calculations~\cite{ON-n3}.  As a
consequence of (\ref{crel}), the value of $s^*$ should be essentially
fixed by the term proportional to $(p^2)^2$ in the inverse propagator,
through the approximate relation
\begin{equation}
s^*-1\simeq c_2.
\label{s4u}
\end{equation}
This equation is satisfied within the precision of our analysis
for $N=0,2$, and well verified by the exact results of the Ising 
model, where $s^*-1-c_2\simeq 10^{-5}$. 

We can conclude that, like models with $N\geq 3$, in the critical
region the two-point Green's function for $N\leq 2$ is almost Gaussian
in a large region around $p^2=0$, i.e., $|p^2/M_G^2|\lesssim 1$, and
the small corrections to Gaussian behavior are essentially determined
by the $(p^2)^2$ term in the expansion of the inverse propagator.

Differences from Gaussian behavior will become important at
sufficiently large momenta, where $G(p)$ should behave as
\begin{equation}
G(p) \sim {1\over p^{2-\eta}}
\end{equation}
where $\eta\neq 0$: $\eta=\case{5}{24}$ for $N=0$,
and $\eta=\case{1}{4}$ for $N=1,2$.

\section{Low-momentum behavior of the Ising model}
\label{Isingmodel}

So far we considered only the critical limit of the two-point Green's
function.  It has however been known for a long time that the
correlation functions of the two-dimensional Ising model can be
computed exactly for arbitrary values of $\beta$. As a consequence we
may in principle check directly our computations for every individual
coordinate-space Green's function.  In practice we may perform our
checks by exploiting a peculiar feature of the square lattice
solution: for sufficiently large values of $\sqrt{x^2+y^2}$ (in units
of the lattice spacing) the asymptotic behavior is described
by~\cite{Cheng-Wu}
\begin{eqnarray}
G(x,y)\simeq &&
\left[ (1-z^2)^2 - 4z^2\right]^{1/4}
(1+z^2)^{1/2} (1-z^2)\times \nonumber \\
&&\int {d\phi_1\over 2\pi}{d\phi_2\over 2\pi}
{e^{i\phi_1 x + i\phi_2 y}
\over (1+z^2)^2 - 2z(1-z^2)(\cos\phi_1+\cos\phi_2)},
\label{eq1}
\end{eqnarray}
where we have introduced the auxiliary variable 
\begin{equation}
z(\beta)={\rm tanh}\beta.
\label{eq2}
\end{equation}
We recognize that the above result (\ref{eq1}) corresponds to the
behavior of a nearest-neighbor quasi-Gaussian model whose
momentum-space propagator has the form
\begin{equation}
\widetilde{G}(p) = {Z(\beta)\over \hat{p}^2+M^2(\beta)}
[1 + g(p,\beta)],
\label{eq3}
\end{equation}
where $g(p,\beta)$ vanishes at the pole $\hat{p}^2 = -M^2(\beta)$,
\begin{equation}
Z(\beta)=\left[ (1-z^2)^2-4z^2\right]^{1/4} {(1+z^2)^{1/2}\over z},
\label{eq4}
\end{equation}
and 
\begin{equation}
M^2(\beta)= {(1+z^2)^2\over z(1-z^2)}-4.
\label{eq5}
\end{equation}
A straightforward but yet unobserved consequence 
of this observation is the algebraic relationship
\begin{equation}
2({\rm cosh}\mu_{\rm s}-1)=
4({\rm cosh}\case{1}{2}\mu_{\rm d}-1)=M^2(\beta),
\label{eq6}
\end{equation}
where $\mu_{\rm s}$ and $\mu_{\rm d}$ are the coefficients of the
long-distance exponential decay (``true mass-gap'') on the side and
along the principal diagonal of the square lattice.  We verified that
Eq.~(\ref{eq6}) is satisfied by our determinations of masses from
wall-wall correlations and is consistent with the known
relationship~\cite{Fisher}
\begin{equation}
\mu_{\rm s}=\ln {\rm coth} \beta-2\beta.
\label{eq7}
\end{equation}
We also checked that the residue at the pole $\hat{p}^2=-M^2$
satisfies Eq.~(\ref{eq4}).

Motivated by this piece of evidence we investigated the possibility
that the asymptotic behavior of the two-point Green's function of the
Ising model on regular two-dimensional lattices be always dictated by
the structure of the propagator
\begin{equation}
\widetilde{G}(p) = {Z(\beta)\over \bar{p}^2+M^2(\beta)}
[1 + g(p,\beta)],
\label{quasi-Gaussian}
\end{equation}
where $\bar{p}^2$ is the massless (nearest-neighbor) Gaussian inverse
propagator appropriate to the lattice at hand, and $g(p,\beta)$
vanishes at the pole $\bar{p}^2 = -M^2(\beta)$.  This conjecture can
be checked by considering the large-distance behavior of the
correlations for the triangular and honeycomb lattices, as a function
of the direction, and comparing the different available mass
definitions with each other and with exact results.

On the triangular lattice we can define a ``true mass-gap'' 
$\mu_{\rm l}$ from the asymptotic behavior of correlations taken along
a straight line of links and a wall-wall inverse correlation length
$\mu_{\rm t}$ evaluated in a direction orthogonal to the above defined
line (see App.~\ref{appb}).  In a Gaussian model one would obtain
\begin{equation}
M_{\rm t}^2 \equiv {8\over 3}
\left({\rm cosh} \case{\sqrt{3}}{2}\mu_{\rm t} -1\right) =
{8\over 3}\left({\rm cosh}\case{1}{2}\mu_{\rm l}-1\right)
\left({\rm cosh}\case{1}{2}\mu_{\rm l}+2\right).
\label{eq8}
\end{equation}
Starting from the known solution~\cite{Stephenson}
\begin{equation}
\mu_{\rm l} = 2\ln\left( \sqrt{1-z+z^2}-\sqrt{z}\right)-2\ln(1-z) -\ln z,
\label{eq9}
\end{equation}
we checked that the relationship (\ref{eq8}) is satisfied, since our
series for $\mu_{\rm t}$ reproduces the expansion of
\begin{equation}
M_{\rm t}^2={2\over 3}
\left[ \left({1+z^2\over 1-z}\right)^2{1\over z}-8 \right].
\label{eq10} 
\end{equation}

Finally on the honeycomb lattice two mutually orthogonal inverse
correlation lengths can be defined by the relationships
\begin{eqnarray}
M_{\rm v}^2&=&{8\over 9} 
\left({\rm cosh} \case{3}{2}\mu_{\rm v}-1 \right), \nonumber \\
M_{\rm h}^2&=&{8\over 3} 
\left({\rm cosh} \case{\sqrt{3}}{2}\mu_{\rm h}-1 \right),
\label{eq11}
\end{eqnarray}
where $\mu_{\rm v}$ and $\mu_{\rm h}$ are defined from the
large-distance exponential behavior respectively of wall-wall
correlation functions $G_{\rm v}^{(w)}(x)$ and $G_{\rm h}^{(w)}(x)$
defined in App.~\ref{appc}.  The Gaussian relationship is
\begin{equation}
M_{\rm v}^2 +2={1\over 8}\left( M_{\rm h}^2+4\right)^2.
\label{eq12}
\end{equation}
Moreover from duality with the weak coupling phase of the triangular
lattice model we obtained
\begin{equation}
\mu_{\rm h}= {1\over \sqrt{3}}
\left[ \ln{\sqrt{ 2{\rm cosh}2\beta-1}-1\over
\sqrt{ 2{\rm cosh}2\beta-1}+1} +2\beta\right]
\label{eq13}
\end{equation}
and we checked that the expansion of the $M_{\rm h}^2$ is consistent
with
\begin{equation}
M_{\rm h}^2= {4\over 3} \left( 2{\rm cosh} 2\beta-1\right)^{1/2}
{\rm coth}\beta-4,
\label{eq14}
\end{equation}
while Eq.~(\ref{eq12}) is satisfied to all known orders of the
strong coupling expansion.

In conclusion we may say that the quasi-Gaussian structure of the
propagator, described by Eq.~(\ref{quasi-Gaussian}), is confirmed for
all regular lattices and is a remarkable piece of evidence in favor of
adopting the quantities $M^2$, $M_{\rm t}^2$ and $M_{\rm h}^2$
respectively as strong-coupling estimators of the mass-gap, sharing
the property of a well-behaved $\beta$-dependence and of a faster
approach to universality in models with quasi-Gaussian behavior.  It
is probably worth observing that, since $g(0,\beta)\neq0$,
Eq.~(\ref{quasi-Gaussian}) does not allow an immediate identification
of the moments $m_{2j}$, and in particular $M_G^2\neq M^2$, and
$Z(\beta)$ is not the standard zero-momentum wave function
renormalization but corresponds to the on-shell definition.

\acknowledgments

We gratefully acknowledge correspondence with R.~Shrock and
conversations with G.~Mussardo, especially concerning bibliography on
the Ising model.

\appendix

\section{Strong-coupling series on the square lattice}
\label{appa}

In order to enable the interested readers to perform their own
analysis, we present most of the series used to derive the results
presented in this paper for $N=0,1,2$. This appendix is devoted to the
square lattice, the following ones to the triangular and honeycomb
lattices.

\subsection{$\protect\bbox{N=0}$}

For the self-avoiding random walk on the square lattice,
longer series of $M_G^2$ can be obtained
from the strong-coupling series of $\chi$ and $m_2$ presented 
in Ref.~\cite{GuttmannO0}. We report our series of
$M_G^2$ for sake of completeness.

\begin{mathletters}
\begin{eqnarray}
M_G^2 &=&
  \beta^{-1} -4 + 3\beta + 2\beta^{3} + 4\beta^{4} - 10\beta^{5} + 
   48\beta^{6} - 128\beta^{7} + 368\beta^{8} - 822\beta^{9} + 
   2008\beta^{10} 
\nonumber \\ &-& 
   4320\beta^{11} + 10336\beta^{12} - 22800\beta^{13} + 
   56312\beta^{14} - 129922\beta^{15} + 327080\beta^{16} - 768414\beta^{17}   
\nonumber \\ &+& 
   1938440\beta^{18} - 4604254\beta^{19} + 
   O(\beta^{20}), \\
M_{\rm s}^2 &=&
  \beta^{-1}-4 + 3\beta + 2\beta^{3} + 4\beta^{4} - 8\beta^{5} + 
   30\beta^{6} - 52\beta^{7} + 140\beta^{8} - 234\beta^{9} + 596\beta^{10}  
\nonumber \\ &-& 
   1010\beta^{11} + 2638\beta^{12} - 4644\beta^{13} + 12634\beta^{14} - 
   23208\beta^{15} +
    O(\beta^{16}), \\
v_4 &=&   \case{1}{16}\beta^{-1} + \case{3}{4} + 
    \case{3}{16}\beta +
   \case{1}{8}\beta^3 + \case{3}{8}\beta^5 - \beta^6 + 
   \case{15}{2}\beta^7 - 19\beta^8 + 
   \case{409}{8}\beta^9 - 103\beta^{10}  
+ \case{511}{2}\beta^{11}\nonumber \\ &-& 539\beta^{12}+
   1468\beta^{13} - 3649\beta^{14} + 
   \case{83211}{8}\beta^{15} - 25668\beta^{16} + 
   \case{534225}{8}\beta^{17} - 154972\beta^{18} \nonumber \\&+& 
   \case{3095629}{8}\beta^{19}+O(\beta^{20}).
\end{eqnarray}
\end{mathletters}

\subsection{$\protect\bbox{N=1}$}

For the Ising model we give strong coupling  
series which cannot be reproduced using known exact results, which
are reported in Sec.~\ref{Isingmodel}.

\begin{mathletters}
\begin{eqnarray}
M_G^2 &=&
  \beta^{-1}-4 + \case{10}{3}\beta + \case{134}{45}\beta^{3} + 
   \case{76}{189}\beta^{5} + \case{19394}{4725}\beta^{7} - 32\beta^{8} + 
   \case{2070328}{18711}\beta^{9} - \case{704}{3}\beta^{10}  
\nonumber \\ &+& 
   \case{233105490328}{638512875}\beta^{11} - \case{20656}{45}\beta^{12} + 
   \case{440148292}{729729}\beta^{13} - \case{256064}{189}\beta^{14} + 
   \case{670306901872438}{162820783125}\beta^{15} - 
   \case{52233344}{4725}\beta^{16}  
\nonumber \\ &+& 
   \case{192016952587260544}{7795859096025}\beta^{17} - 
   \case{4476104704}{93555}\beta^{18} + 
   \case{133522860364557505628}{1531329465290625}\beta^{19} + 
   O(\beta^{20}), \\
v_4 &=&   \case{1}{16}\beta^{-1}+ \case{3}{4} + \case{5}{24}\beta +
   \case{67}{360}\beta^3 + \case{19}{756}\beta^5+
   \case{9697}{37800}\beta^7 + 2\beta^8 - 
   \case{339961}{37422}\beta^9 + \case{44}{3}\beta^{10}+
   \case{8705774291}{1277025750}\beta^{11}\nonumber \\&-&
   \case{4514}{45}\beta^{12}+
   \case{3986722469}{14594580}\beta^{13} - 
   \case{68668}{189}\beta^{14}-
   \case{115832206185781}{1302566265000}\beta^{15} +
   \case{6752894}{4725}\beta^{16} \nonumber \\&-& 
   \case{21607992820912952}{7795859096025}\beta^{17} + 
   \case{34100716}{93555}\beta^{18} + 
   \case{71772260149691061407}{6125317861162500}\beta^{19}+O(\beta^{20}).
\end{eqnarray}
\end{mathletters}

\subsection{$\protect\bbox{N=2}$}

\begin{mathletters}
\begin{eqnarray}
E &=&
  \beta  + \case{3}{2}\beta^{3} + \case{1}{3}\beta^{5} - 
   \case{31}{48}\beta^{7} - \case{731}{120}\beta^{9} - 
   \case{29239}{1440}\beta^{11} - \case{265427}{5040}\beta^{13} - 
   \case{75180487}{645120}\beta^{15} - 
   \case{6506950039}{26127360}\beta^{17}  
\nonumber \\ &-& 
   \case{1102473407093}{2612736000}\beta^{19} - 
   \case{6986191770643}{14370048000}\beta^{21} +
   O(\beta^{23}), \\
\chi &=&
  1 + 4\beta + 12\beta^{2} + 34\beta^{3} + 88\beta^{4} + 
   \case{658}{3}\beta^{5} + 529\beta^{6} + \case{14933}{12}\beta^{7} + 
   \case{5737}{2}\beta^{8} + \case{389393}{60}\beta^{9}  
\nonumber \\ &+& 
   \case{2608499}{180}\beta^{10} + \case{3834323}{120}\beta^{11} + 
   \case{1254799}{18}\beta^{12} + \case{84375807}{560}\beta^{13} + 
   \case{6511729891}{20160}\beta^{14} + \case{66498259799}{96768}\beta^{15}   
\nonumber \\ &+& 
   \case{1054178743699}{725760}\beta^{16} + 
   \case{39863505993331}{13063680}\beta^{17} + 
   \case{19830277603399}{3110400}\beta^{18} + 
   \case{8656980509809027}{653184000}\beta^{19}  
\nonumber \\ &+& 
   \case{2985467351081077}{108864000}\beta^{20} + 
   \case{811927408684296587}{14370048000}\beta^{21} + 
   O(\beta^{22}), \\
M_G^2 &=&
  \beta^{-1}-4 + \case{7}{2}\beta + \case{41}{12}\beta^{3} - 
   {{\beta }^4} + \case{15}{16}\beta^{5} - \case{25}{3}\beta^{6} + 
   \case{9491}{720}\beta^{7} - \case{431}{9}\beta^{8} + 
   \case{206411}{2880}\beta^{9} - \case{17803}{360}\beta^{10}  
\nonumber \\ &-& 
   \case{41122019}{241920}\beta^{11} + \case{876403}{1728}\beta^{12} - 
   \case{1413373319}{1935360}\beta^{13} - \case{15006841}{181440}\beta^{14} + 
   \case{337093786457}{130636800}\beta^{15} - 
   \case{4777620367}{1036800}\beta^{16}   
\nonumber \\ &+& 
   \case{17847363647}{1741824000}\beta^{17} + 
   \case{68513340691}{3732480}\beta^{18} - 
   \case{16133717627082721}{344881152000}\beta^{19} + 
   O(\beta^{20}), \\
M_{\rm s}^2 &=&
  \beta^{-1}-4 + \case{7}{2}\beta + \case{41}{12}\beta^{3} - 
   {{\beta }^4} + \case{7}{16}\beta^{5} - \case{29}{6}\beta^{6} + 
   \case{281}{720}\beta^{7} - \case{193}{18}\beta^{8} - 
   \case{149}{2880}\beta^{9} - \case{5141}{720}\beta^{10}  
\nonumber \\ &-& 
   \case{6120227}{241920}\beta^{11} + \case{24907}{540}\beta^{12} - 
   \case{788579333}{5806080}\beta^{13} + \case{95728039}{362880}\beta^{14} - 
   \case{63069969313}{130636800}\beta^{15} +
    O(\beta^{16}), \\
v_4 &=&   \case{1}{16}\beta^{-1} + {3\over 4} + \case{7}{32}\beta+
   \case{41}{192}\beta^3 - \case{49}{256}\beta^5+
   \case{1}{4}\beta^6 - \case{8749}{11520}\beta^7 + 
   \case{67}{48}\beta^8 - \case{122549}{46080}\beta^9 - 
   \case{2153}{144}\beta^{10}\nonumber \\&+&
   \case{249335197}{3870720}\beta^{11}-
   \case{40951}{320}\beta^{12}+
   \case{1389732217}{30965760}\beta^{13} + 
   \case{55582271}{103680}\beta^{14} - 
   \case{3706449404743}{2090188800}\beta^{15} + 
   \case{6252985429}{2903040}\beta^{16} \nonumber \\ &+& 
   \case{75252500337407}{27869184000}\beta^{17}- 
   \case{202521546511}{12441600}\beta^{18} + 
   \case{163636654204247999}{5518098432000}\beta^{19}+O(\beta^{20}).
\end{eqnarray}
\end{mathletters}

\section{Strong-coupling series on the triangular lattice}
\label{appb}

The sites $\vec{x}$ of a finite periodic triangular 
lattice can be represented in Cartesian coordinates by
\begin{eqnarray}
&&\vec{x}(l_1,l_2)= l_1\vec{\eta}_1 
+ l_2\vec{\eta}_2,\nonumber \\
&& l_1=1,...L_1, \ l_2=1,...L_2,\nonumber \\
&& \vec{\eta}_1=\left( 1,0\right),\quad
\vec{\eta}_2= \left( {1\over 2},{\sqrt{3}\over 2}\right).
\label{trcoord}
\end{eqnarray}
In order to define a mass-gap estimator,
one may consider the following  wall-wall correlation
function 
\begin{equation}
G_{\rm t}^{(\rm w)}\left({\sqrt{3}\over 2}l_2\right)=
\sum_{l_1} G(l_1\vec{\eta}_1+l_2\vec{\eta}_2).
\label{walldeftr}
\end{equation}
An estimator $\mu_{\rm t}$ of the mass-gap can be extracted from the
long distance behavior of $G_{\rm t}^{(\rm w)}(x)$,
indeed for $x\gg 1$
\begin{equation}
G_{\rm t}^{(\rm w)}(x)\propto e^{-\mu_{\rm t} x}.
\label{ldgtr}
\end{equation}
In view of a strong-coupling analysis, it is convenient to use another
estimator of the mass-gap derived from $\mu_{\rm t}$:
\begin{equation}
M_{\rm t}^2\equiv {8\over 3}
\left( {\rm cosh} {\sqrt{3}\over 2} \mu_{\rm t} - 1\right).
\label{Mt}
\end{equation}
More details can be found in Ref.~\cite{ON-n3}.

In the following we show, for $N=0,1,2$, some of the strong-coupling
series used in the analysis of the ${\rm O}(N)$ $\sigma$ models on the
triangular lattice presented in this paper.

\subsection{$\protect\bbox{N=0}$}

For the self-avoiding random walk on the triangular lattice,
longer series of $M_G^2$ can be obtained
from the strong-coupling series of $\chi$ and $m_2$ presented 
in Ref.~\cite{GuttmannO0t}.

\begin{mathletters}
\begin{eqnarray}
M_G^2 &=&
   \case{2}{3}\beta^{-1} -4 + \case{10}{3}\beta + 4\beta^{2} + 
   \case{16}{3}\beta^{3} + \case{40}{3}\beta^{4} + \case{88}{3}\beta^{5} + 
   \case{88}{3}\beta^{6} + 228\beta^{7} + \case{1808}{3}\beta^{8} + 
   \case{4352}{3}\beta^{9}
\nonumber \\ &+& 
   \case{18356}{3}\beta^{10} + 
   \case{52792}{3}\beta^{11} + 60540\beta^{12} + \case{631184}{3}\beta^{13} +
   O(\beta^{14}), \\
M_{\rm t}^2 &=&
    \case{2}{3}\beta^{-1} -4 + \case{10}{3}\beta + 4\beta^{2} + 
   \case{17}{3}\beta^{3} + \case{35}{3}\beta^{4} + \case{47}{2}\beta^{5} + 
   \case{205}{3}\beta^{6} + 188\beta^{7} + \case{2213}{4}\beta^{8} + 
   \case{41909}{24}\beta^{9}
\nonumber \\ &+& 
   \case{33181}{6}\beta^{10} +
   O(\beta^{11}), \\
v_4 &=&  \case{1}{24}\beta^{-1} +\case{3}{4}+ \case{5}{24}\beta+
   \case{1}{4}\beta^2 + \case{1}{3}\beta^3+
   \case{1}{3}\beta^4 + \case{4}{3}\beta^5+
   \case{59}{6}\beta^6 + \case{55}{4}\beta^7+
   \case{98}{3}\beta^8 + \case{1135}{6}\beta^9\nonumber \\&+&
   \case{4529}{12}\beta^{10}
   + \case{4783}{3}\beta^{11}+
   \case{21295}{4}\beta^{12}+ \case{100165}{6}\beta^{13}+O(\beta^{14}).
\end{eqnarray}
\end{mathletters}

\subsection{$\protect\bbox{N=1}$}

For the Ising model we give strong coupling  
series which cannot be reproduced using known exact results.

\begin{mathletters}
\begin{eqnarray}
M_G^2 &=&
   \case{2}{3}\beta^{-1} -4 + \case{32}{9}\beta + \case{16}{3}\beta^{2} + 
   \case{928}{135}\beta^{3} + \case{64}{9}\beta^{4} + 
   \case{23944}{2835}\beta^{5} - \case{1648}{135}\beta^{6} + 
   \case{5008}{14175}\beta^{7} + \case{106864}{945}\beta^{8}  
\nonumber \\ &+& 
   \case{6459424}{280665}\beta^{9} - \case{18680128}{42525}\beta^{10} - 
   \case{200433692584}{1915538625}\beta^{11} + 
   \case{2151999728}{1403325}\beta^{12} + 
   \case{35136345008}{54729675}\beta^{13} +
   O(\beta^{14}), 
\nonumber \\ \\
v_4 &=&   \case{1}{24}\beta^{-1} + \case{3}{4} + \case{2}{9}\beta+
   \case{1}{3}\beta^2 + \case{58}{135}\beta^3+
   \case{4}{9}\beta^4 + \case{2993}{5670}\beta^5 + 
   \case{437}{135}\beta^6 + \case{313}{14175}\beta^7 - 
   \case{19781}{945} \beta^8 \nonumber \\ &+&
   \case{403714}{280665}\beta^9 +
   \case{3986522}{42525}\beta^{10} - 
   \case{354526855073}{3831077250}\beta^{11} - 
   \case{960767707}{1403325}\beta^{12} - 
   \case{18090444637}{54729675}\beta^{13}+O(\beta^{14}).
\end{eqnarray}
\end{mathletters}

\subsection{$\protect\bbox{N=2}$}

\begin{mathletters}
\begin{eqnarray}
E &=&
  \beta  + 2\beta^{2} + \case{7}{2}\beta^{3} + 5\beta^{4} + 
   \case{35}{6}\beta^{5} + \case{14}{3}\beta^{6} - \case{81}{16}\beta^{7} - 
   \case{3769}{72}\beta^{8} - \case{165161}{720}\beta^{9} - 
   \case{7821}{10}\beta^{10}
\nonumber \\ &-& 
   \case{20160371}{8640}\beta^{11} - 
   \case{27984359}{4320}\beta^{12} - \case{87289819}{5040}\beta^{13} - 
   \case{10256893919}{226800}\beta^{14} - 
   \case{3357272555039}{29030400}\beta^{15} +
   O(\beta^{16}), 
\nonumber \\ \\
\chi &=&
  1 + 6\beta + 30\beta^{2} + 135\beta^{3} + 570\beta^{4} + 2306\beta^{5} + 
   \case{18083}{2}\beta^{6} + \case{276657}{8}\beta^{7} + 
   \case{777805}{6}\beta^{8}
\nonumber \\ &+& 
   \case{14339641}{30}\beta^{9} + 
   \case{208590287}{120}\beta^{10} + \case{8995595389}{1440}\beta^{11} + 
   \case{3199713875}{144}\beta^{12} + \case{65793037351}{840}\beta^{13}  
\nonumber \\ &+& 
   \case{165647319078571}{604800}\beta^{14} + 
   \case{4600845479023849}{4838400}\beta^{15} + 
   O(\beta^{16}), \\
M_G^2 &=&
   \case{2}{3}\beta^{-1} -4 + \case{11}{3}\beta + 6\beta^{2} + 
   \case{143}{18}\beta^{3} + \case{46}{9}\beta^{4} - 
   \case{391}{72}\beta^{5} - \case{5219}{108}\beta^{6} - 
   \case{5296}{45}\beta^{7} - \case{33287}{180}\beta^{8}  
\nonumber \\ &-& 
   \case{679729}{1296}\beta^{9} - \case{2052143}{1080}\beta^{10} - 
   \case{1436935039}{362880}\beta^{11} - 
   \case{4952351659}{1360800}\beta^{12} - 
   \case{87992319949}{43545600}\beta^{13} +
   O(\beta^{14}) \\
M_{\rm t}^2 &=&
   \case{2}{3}\beta^{-1} -4 + \case{11}{3}\beta + 6\beta^{2} + 
   \case{283}{36}\beta^{3} + \case{49}{9}\beta^{4} - \case{53}{8}\beta^{5} - 
   \case{8425}{216}\beta^{6} - \case{990757}{8640}\beta^{7} - 
   \case{45549}{160}\beta^{8}
\nonumber \\ &-& 
   \case{16833083}{25920}\beta^{9} - 
   \case{35865709}{25920}\beta^{10} +
   O(\beta^{11}), \\
v_4 &=&  \case{1}{24}\beta^{-1} + \case{3}{4} + \case{11}{48}\beta+
   \case{3}{8}\beta^2 + \case{143}{288}\beta^3+
   \case{4}{9}\beta^4- \case{103}{1152}\beta^5-
   \case{827}{1728}\beta^6 - \case{6271}{720}\beta^7-
   \case{39709}{960}\beta^8 \nonumber \\ &-& 
   \case{1243813}{20736}\beta^9 +
    \case{85031}{3456}\beta^{10} 
- \case{741356239}{5806080}\beta^{11} - 
   \case{7581779911}{5443200}\beta^{12} - 
   \case{1477616543629}{696729600}\beta^{13}+O(\beta^{14}).
\end{eqnarray}
\end{mathletters}

\section{Strong-coupling series on the honeycomb lattice}
\label{appc}

The sites $\vec{x}$ of a finite periodic honeycomb 
lattice can be represented in Cartesian coordinates by
\begin{eqnarray}
&&\vec{x}= \vec{x}\,'+p\,\vec{\eta}_p
\nonumber \\
&&\vec{x}\,'= l_1\,\vec{\eta}_1 
+ l_2\,\vec{\eta}_2,\nonumber \\
&& l_1=1,...L_1, \ l_2=1,...L_2, \ 
p=0,1,\nonumber \\
&& \vec{\eta}_1=\left( {3\over 2},{\sqrt{3}\over 2}
\right),\quad
\vec{\eta}_2= \left( 0,\sqrt{3}\right),
\quad \vec{\eta}_p=\left(1,0\right).
\label{hocoord}
\end{eqnarray}
In order to define a mass-gap estimator,
one may consider the following  wall-wall correlation
functions
\begin{equation}
G^{(\rm w)}_{\rm v}(\case{3}{2}l_1)=\sum_{l_2} 
G(l_1\vec{\eta}_1+l_2\vec{\eta}_2),
\label{g1}
\end{equation}
with the sum running over sites of positive parity forming a vertical
line; 
\begin{equation}
G^{(\rm w)}_{\rm h}(\case{1}{2} \sqrt{3} l) =
\sum_{l_2,p} 
G\bigl((l-2l_2)\vec{\eta}_1+l_2\vec{\eta}_2+p\vec{\eta}_p\bigr), 
\label{g2}
\end{equation}
where the sum is performed over all sites having the same coordinate
$x_2$.

Estimators $\mu_{\rm v}$ and $\mu_{\rm h}$ of the mass-gap can be
extracted from the long distance behavior respectively of $G_{\rm
v}^{(\rm w)}(x)$ and $G_{\rm h}^{(\rm w)}(x)$, indeed for $x\gg 1$
\begin{eqnarray}
G_{\rm h}^{(\rm w)}(x)&\propto& e^{-\mu_{\rm v} x},\nonumber \\
G_{\rm h}^{(\rm w)}(x)&\propto& e^{-\mu_{\rm h} x}.
\label{ldgho}
\end{eqnarray}
In view of a strong-coupling analysis, it is convenient to use
the following estimators of
the mass-gap derived from $\mu_{\rm v}$ and $\mu_{\rm h}$:
\begin{eqnarray}
M_{\rm v}^2&\equiv& {8\over 9}
\left( {\rm cosh} \case{3}{2} \mu_{\rm h} - 1\right),\nonumber \\ 
M_{\rm h}^2&\equiv &{8\over 3}
\left( {\rm cosh} \case{\sqrt{3}}{2} \mu_{\rm h} - 1\right).
\label{Mh}
\end{eqnarray}
More details can be found in Ref.~\cite{ON-n3}.

In the following we show, for $N=0,1,2$, some of the strong-coupling
series used in the analysis of ${\rm O}(N)$ $\sigma$ models on the
honeycomb lattice presented in this paper.

\subsection{$\protect\bbox{N=0}$}

For the self-avoiding random walk on the honeycomb lattice,
longer series of $M_G^2$ can be obtained
from the strong-coupling series of $\chi$ and $m_2$ presented 
in Ref.~\cite{GuttmannO0}.

\begin{mathletters}
\begin{eqnarray}
M_G^2 &=&
   \case{4}{3}\beta^{-1} -4 + \case{8}{3}\beta + 8\beta^{6} - 
   \case{56}{3}\beta^{7} + 24\beta^{8} - 32\beta^{9} + 96\beta^{10} - 
   \case{656}{3}\beta^{11} + 320\beta^{12} - 296\beta^{13}
\nonumber \\ &+& 
   416\beta^{14} - 
   1192\beta^{15} + 2848\beta^{16} - \case{13304}{3}\beta^{17} + 
   5768\beta^{18} - \case{27664}{3}\beta^{19} + 20024\beta^{20} - 
   38520\beta^{21}
\nonumber \\ &+& 
   63368\beta^{22} - 100104\beta^{23} + 183352\beta^{24} - 
   \case{1039744}{3}\beta^{25} + 621096\beta^{26} - 
   \case{3093176}{3}\beta^{27}
\nonumber \\ &+& 
   1791168\beta^{28} +
    O(\beta^{29}), \\
M_{\rm h}^2 &=&
   \case{4}{3}\beta^{-1} -4 + \case{8}{3}\beta + \case{4}{3}\beta^{5} + 
   \case{8}{3}\beta^{9} + \case{8}{3}\beta^{10} - 4\beta^{11} + 8\beta^{12} + 
   \case{16}{3}\beta^{13} + 8\beta^{14} + \case{8}{3}\beta^{15} + 
   16\beta^{16}
\nonumber \\ &+& 
   \case{148}{3}\beta^{17} + 36\beta^{18} + 
   \case{176}{3}\beta^{19} + \case{100}{3}\beta^{20} + 
   \case{1532}{3}\beta^{21} - 248\beta^{22} + \case{4348}{3}\beta^{23} - 
   \case{3184}{3}\beta^{24} +
   O(\beta^{25}), 
\nonumber \\ 
v_4 &=&  \case{1}{12}\beta^{-1}+\case{3}{4}+ \case{1}{6}\beta -
   \case{1}{2}\beta^6 + \case{11}{6}\beta^7-
   \case{3}{2}\beta^8 + \beta^9 - 6\beta^{10} + 
   \case{70}{3}\beta^{11} - 35\beta^{12}+
   \case{11}{2}\beta^{13} +  49\beta^{14} \nonumber \\ &+& 
   \case{75}{2}\beta^{15} - 397\beta^{16} + 
   \case{4631}{6}\beta^{17} - \case{1051}{2}\beta^{18}-
   \case{349}{3}\beta^{19}- \case{1087}{2}\beta^{20}+
   \case{8001}{2}\beta^{21} - \case{15883}{2}\beta^{22} +
   \case{17989}{2}\beta^{23}\nonumber \\&-&  
   \case{19691}{2}\beta^{24}+
   \case{72668}{3}\beta^{25} - \case{112563}{2}\beta^{26}+
   \case{536981}{6}\beta^{27} - 110355\beta^{28}+O(\beta^{29}).
\end{eqnarray}
\end{mathletters}

\subsection{$\protect\bbox{N=1}$}

\begin{mathletters}
\begin{eqnarray}
M_G^2 &=&
   \case{4}{3}\beta^{-1} -4 + \case{28}{9}\beta - \case{124}{135}\beta^{3} + 
   \case{8576}{2835}\beta^{5} - \case{14692}{2025}\beta^{7} + 
   \case{5338616}{280665}\beta^{9} - 
   \case{90947891648}{1915538625}\beta^{11} - 32\beta^{12}   
\nonumber \\ &+& 
   \case{1583805616}{7818525}\beta^{13} + 96\beta^{14} - 
   \case{406965884456828}{488462349375}\beta^{15} - 
   \case{608}{15}\beta^{16} + \case{360870502928894432}{116937886440375
    }\beta^{17} - \case{1219136}{945}\beta^{18}   
\nonumber \\ &-& 
   \case{6493740451647884584}{656284056553125}\beta^{19} + 
   \case{624064}{75}\beta^{20} + 
   \case{1179228814388026215376}{40343570821929375}\beta^{21} - 
   \case{5843265248}{155925}\beta^{22}  
\nonumber \\ &-& 
   \case{48910471162936574893856768}{605758715891269565625}\beta^{23} + 
   \case{405278723648}{2837835}\beta^{24} + 
   \case{1024764052182397586576416}{4754777989727390625}\beta^{25}  
\nonumber \\ &-& 
   \case{1697456183968}{3378375}\beta^{26} - 
   \case{950155935558179228231150591072}{1693960980510228821015625}\beta^{27
    } + \case{54851554589151328}{32564156625}\beta^{28} +
    O(\beta^{29}), \\
v_4 &=&   \case{1}{12}\beta^{-1}+\case{3}{4} + 
   \case{7}{36}\beta - 
   \case{31}{540}\beta^3 + 
   \case{536}{2835}\beta^5 - 
   \case{3673}{8100}\beta^7 + 
   \case{667327}{561330}\beta^9 - 
   \case{5684243228}{1915538625}\beta^{11} + 
   2\beta^{12}\nonumber \\ &+& 
   \case{5165551}{7818525}\beta^{13} - 
   6\beta^{14} + 
   \case{93643468635793}{1953849397500}\beta^{15} - 
   \case{1042}{15}\beta^{16} - 
   \case{27120807726815398}{116937886440375}\beta^{17} + 
   \case{643196}{945}\beta^{18} \nonumber \\ &+& 
   \case{86312750362752061}{187509730443750}\beta^{19} - 
   \case{258424}{75}\beta^{20} + 
   \case{33629885325845121661}{40343570821929375}\beta^{21} + 
   \case{1973399678}{155925}\beta^{22} \nonumber \\ &-& 
   \case{6840786318771414403278548}{605758715891269565625}\beta^{23} - 
   \case{106606055168}{2837835}\beta^{24} + 
   \case{282271419983204843625526}{4754777989727390625}\beta^{25} + 
   \case{44788447114}{482625}\beta^{26} \nonumber \\ &-& 
   \case{402797762032926523234583974442}{1693960980510228821015625}\beta^{27} - 
   \case{5992229992104838}{32564156625}\beta^{28}
+O(\beta^{29})
\end{eqnarray}
\end{mathletters}

\subsection{$\protect\bbox{N=2}$}

\begin{mathletters}
\begin{eqnarray}
E &=&
  \beta  - \case{1}{2}\beta^{3} + \case{7}{3}\beta^{5} - 
   \case{395}{48}\beta^{7} + \case{1173}{40}\beta^{9} - 
   \case{473243}{4320}\beta^{11} + \case{6293627}{15120}\beta^{13} - 
   \case{346093553}{215040}\beta^{15}   
\nonumber \\ &+& 
   \case{23497364693}{3732480}\beta^{17} - 
   \case{64962730739719}{2612736000}\beta^{19} + 
   \case{474090720713083}{4790016000}\beta^{21} - 
   \case{1641257090013388013}{4138573824000}\beta^{23}   
\nonumber \\ &+& 
   \case{42984420336380838389}{26900729856000}\beta^{25} - 
   \case{11369733294965786406529}{1757514350592000}\beta^{27} + 
   \case{1733398746685522588378351}{65906788147200000}\beta^{29} +
   O(\beta^{30}), 
\nonumber \\ \\
\chi &=&
  1 + 3\beta + 6\beta^{2} + \case{21}{2}\beta^{3} + 18\beta^{4} + 
   31\beta^{5} + \case{95}{2}\beta^{6} + \case{1045}{16}\beta^{7} + 
   \case{403}{4}\beta^{8} + \case{6919}{40}\beta^{9} + 
   \case{14149}{60}\beta^{10}
\nonumber \\ &+& 
   \case{68273}{288}\beta^{11} + 
   \case{138307}{360}\beta^{12} + \case{9157051}{10080}\beta^{13} + 
   \case{42124273}{40320}\beta^{14} - \case{13183321}{645120}\beta^{15} + 
   \case{130286011}{161280}\beta^{16}   
\nonumber \\ &+& 
   \case{58701184637}{8709120}\beta^{17} + 
   \case{246444397309}{43545600}\beta^{18} - 
   \case{12790078293739}{870912000}\beta^{19} - 
   \case{79551567889}{13608000}\beta^{20} + 
   \case{154021837152677}{1916006400}\beta^{21}  
\nonumber \\ &+& 
   \case{1452164594591761}{28740096000}\beta^{22} - 
   \case{393634368786168197}{1379524608000}\beta^{23} - 
   \case{4660955848386121}{31352832000}\beta^{24} + 
   \case{10915691174925870017}{8966909952000}\beta^{25}  
\nonumber \\ &+& 
   \case{41989331871750076949}{62768369664000}\beta^{26} - 
   \case{8481318776641386327367}{1757514350592000}\beta^{27} - 
   \case{828979117543657737823}{329533940736000}\beta^{28}  
\nonumber \\ &+& 
   \case{5226218120804763962092657}{263627152588800000}\beta^{29} + 
   \case{21701722199756349611186159}{2109017220710400000}\beta^{30} + 
   O(\beta^{31}), \\
M_G^2 &=&
   \case{4}{3}\beta^{-1} -4 + \case{10}{3}\beta - \case{13}{9}\beta^{3} + 
   \case{59}{12}\beta^{5} - 2\beta^{6} - \case{3347}{270}\beta^{7} + 
   \case{35}{6}\beta^{8} + \case{238009}{6480}\beta^{9} - 
   \case{493}{18}\beta^{10}
\nonumber \\ &-& 
   \case{19392227}{181440}\beta^{11} + 
   \case{4388}{45}\beta^{12} + \case{1467214247}{4354560}\beta^{13} - 
   \case{3846767}{12960}\beta^{14} - 
   \case{245879581721}{195955200}\beta^{15} + 
   \case{362651221}{362880}\beta^{16}  
\nonumber \\ &+& 
   \case{6669774367471}{1306368000}\beta^{17} - 
   \case{45385487873}{10886400}\beta^{18} - 
   \case{5401824824719549}{258660864000}\beta^{19} + 
   \case{68257961593}{3483648}\beta^{20}  
\nonumber \\ &+& 
   \case{258003704533726433}{3103930368000}\beta^{21} - 
   \case{362403210060397}{3919104000}\beta^{22} - 
   \case{5054778739819764833}{15692092416000}\beta^{23} + 
   \case{72793501494263779}{172440576000}\beta^{24}  
\nonumber \\ &+& 
   \case{290687412809274634879279}{237264437329920000}\beta^{25} - 
   \case{202156253372553206149}{108637562880000}\beta^{26} - 
   \case{43738864245549216552954097}{9490577493196800000}\beta^{27}  
\nonumber \\ &+& 
   \case{45054828678355702664561}{5649153269760000}\beta^{28} +
    O(\beta^{29}), \\
M_{\rm h}^2 &=&
   \case{4}{3}\beta^{-1} -4 + \case{10}{3}\beta - \case{13}{9}\beta^{3} + 
   \case{55}{12}\beta^{5} - \case{1}{3}\beta^{6} - 
   \case{7429}{540}\beta^{7} - \case{2}{9}\beta^{8} + 
   \case{282139}{6480}\beta^{9} - \case{43}{216}\beta^{10}  
\nonumber \\ &-& 
   \case{26145491}{181440}\beta^{11} + \case{613}{540}\beta^{12} + 
   \case{2158358071}{4354560}\beta^{13} - \case{224587}{77760}\beta^{14} - 
   \case{344839817111}{195955200}\beta^{15} + 
   \case{1698299}{272160}\beta^{16}  
\nonumber \\ &+& 
   \case{8350838655511}{1306368000}\beta^{17} + 
   \case{63590671}{130636800}\beta^{18} - 
   \case{3061458683224637}{129330432000}\beta^{19} - 
   \case{1894590323}{52254720}\beta^{20} 
\nonumber \\ &+& 
   \case{39427276163585267}{443418624000}\beta^{21} + 
   \case{5154851721889}{23514624000}\beta^{22} - 
   \case{5303030533425785401}{15692092416000}\beta^{23} - 
   \case{174214610003233}{147806208000}\beta^{24} 
\nonumber \\ &+& 
   O(\beta^{25}), \\
v_4 &=&  \case{1}{12}\beta^{-1} + \case{3}{4} + 
   \case{5}{24}\beta - 
   \case{13}{144}\beta^3 + 
   \case{59}{192}\beta^5 + 
   \case{1}{8}\beta^6 - 
   \case{5507}{4320}\beta^7 - 
   \case{71}{96}\beta^8 + 
   \case{495049}{103680}\beta^9 + 
   \case{961}{288}\beta^{10}\nonumber \\ &-& 
   \case{47837987}{2903040}\beta^{11} - 
   \case{43397}{2880}\beta^{12} + 
   \case{4159541927}{69672960}\beta^{13} + 
   \case{11139143}{207360}\beta^{14} - 
   \case{643769125241}{3135283200}\beta^{15} - 
   \case{1093077841}{5806080}\beta^{16} \nonumber \\ &+& 
   \case{13258559750671}{20901888000}\beta^{17} + 
   \case{148926884453}{174182400}\beta^{18} - 
   \case{1148899892904427}{591224832000}\beta^{19} - 
   \case{6289014713053}{1393459200}\beta^{20}  
\nonumber \\ &+& 
   \case{353359211049440273}{49662885888000}\beta^{21} +
   \case{202048769925301}{8957952000}\beta^{22} - 
   \case{2704447377391331531}{83691159552000}\beta^{23} - 
   \case{278880464307951691}{2759049216000}\beta^{24} 
\nonumber \\ &+& 
   \case{54781761414365119518029}{345111908843520000}\beta^{25} +
   \case{715504863884795060149}{1738201006080000}\beta^{26} - 
   \case{10548822433407426077233907}{13804476353740800000}\beta^{27} 
\nonumber \\ &-& 
   \case{140706622546312581163859}{90386452316160000}\beta^{28}
+ O(\beta^{29}).
\end{eqnarray}
\end{mathletters}



\input psfig
\pssilent
\def\centerline#1{\hbox to \hsize {\hss #1\hss}}

\begin{figure}
\centerline{\psfig{figure=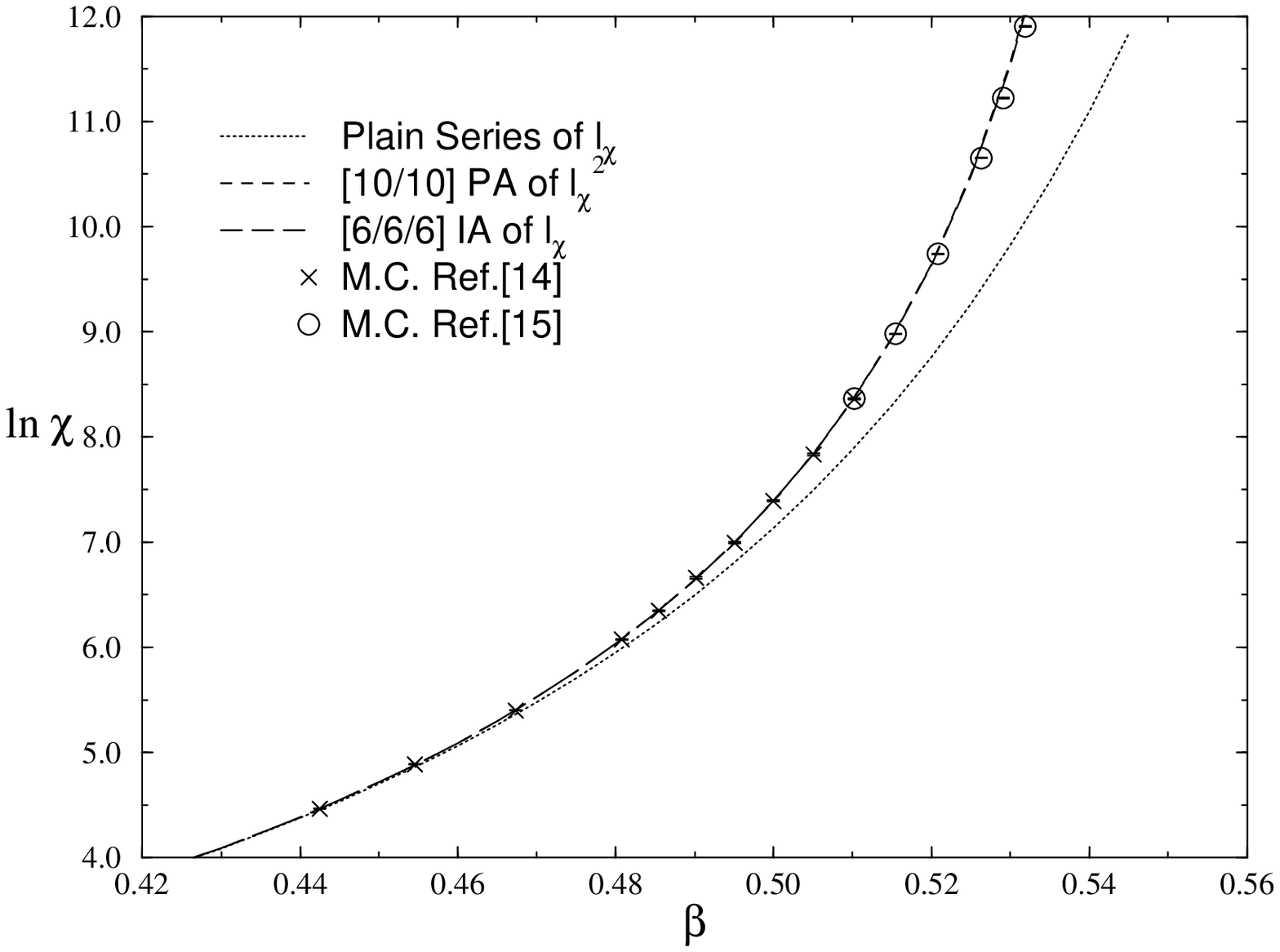}}
\caption{$\ln\chi$ vs.\ $\beta$. Beside Monte Carlo data from 
Refs.~\protect\cite{Gupta,Kim}, we show curves constructed from the
plain series of $\ln\chi$, from the [10/10] PA of $l_\chi^2$, and from
the [6/6/6] IA of $l_\chi$ biased at $\beta_c=0.5583$, such that
$\sigma\simeq 0.50$.}
\label{figchi}
\end{figure}


\begin{table}
\squeezetable
\caption{
For various values of $N<2$ and for all lattices considered we report
$\beta_c^{(\chi)}$ and $\gamma$ as obtained from a Dlog-PA analysis of
the strong-coupling series of $\chi$, and $\beta_c^{(\xi)}$ and $\nu$
from that of $\xi^2_G$.  Defective Dlog-PA's, i.e., those with
spurious singularities close to the real axis for ${\rm Re} \beta
\protect\lesssim \beta_c$ (e.g., ${\rm Re} \beta < 1.1 \beta_c$) are
discarded. An asterisk indicates that most of Dlog-PA's considered are
defective and the estimate comes just from a few of them, or in the
cases where numbers are not shown that all Dlog-PA's are defective, so
that no estimate can be extracted.
\label{Nm2tab}}
\begin{tabular}{ccr@{}lr@{}lr@{}lr@{}l}
\multicolumn{1}{c}{$N$}&
\multicolumn{1}{c}{lattice}&
\multicolumn{2}{c}{$\beta_c^{(\chi)}$}& 
\multicolumn{2}{c}{$\gamma$}& 
\multicolumn{2}{c}{$\beta_c^{(\xi)}$}& 
\multicolumn{2}{c}{$\nu$}\\
\tableline \hline
$-\case{7}{4}$& triangular & 0&.1728(4)   & 1&.04(2)   & *& & *&  \\ 
    & Eq.~(\protect\ref{nm2expf}) &  &  & 1&.05371...&  &         & 0&.547925... \\
\hline
$-\case{3}{2}$& square     & 0&.258(1)    & 0&.80(3)   & 0&.252(1)  & 0&.65(5)  \\ 
    & triangular & 0&.1875(2)   & 1&.09(1)   & 0&.1888(4) & 0&.64(1)  \\ 
    & honeycomb  &*0&.339(1)    &*0&.80(5)   & 0&.3319(1) & 0&.52(1)  \\ 
    & Eq.~(\protect\ref{nm2expf})  &  & & 1&.08759...&  &         & 0&.574690... \\
\hline
-1  & square     &  0&.3144(1)  & 1&.13(1)   & *0&.3141   &  *0&.60  \\
    & triangular &  0&.2082(1)  & 1&.14(1)   & 0&.2086(1) & 0&.642(3)  \\
    & honeycomb  &  0&.42332(5) & 1&.11(1)   & 0&.4240(6) & 0&.64(4)  \\
    & Eq.~(\protect\ref{nm2expf})   & & & 1&.15625   &     &      & 0&.625  \\
\hline
$-\case{1}{2}$& square     &  0&.34919(6) & 1&.233(5)  & 0&.34922(2)& 0&.681(1)  \\
    & triangular &  0&.2252(2)  & 1&.23(2)   & 0&.22528(2)& 0&.687(1)  \\
    & honeycomb  &  0&.48504(3) & 1&.233(3)  & 0&.48498(2)& 0&.672(1)  \\
    & Eq.~(\protect\ref{nm2expf})   & & & 1&.23758...&    &       & 0&.680715... \\
\hline
0   & square     &  0&.37900(4) & 1&.334(2)  & 0&.37905(2)& 0&.750(1) \\
    & triangular &  0&.24087(4) & 1&.332(5)  & 0&.24092(3)& 0&.750(2)  \\
    & honeycomb  &  0&.54117(3) & 1&.341(3)  & 0&.54116(1)& 0&.748(1)  \\
    & Eq.~(\protect\ref{nm2expf})   &  && 1&.34375   &     &      & 0&.75 \\
\hline
$\case{1}{2} $& square     &  0&.40854(1) & 1&.494(1)  &0&.408530(3)& 0&.8453(2)  \\
    & triangular &  0&.25686(5) & 1&.484(4)  &0&.25692(1) & 0&.8450(1)  \\
    & honeycomb  &  0&.59730(2) & 1&.492(1)  &0&.59731(1) & 0&.8446(1)  \\
    & Eq.~(\protect\ref{nm2expf})   &  && 1&.49641...&     &      & 0&.845852... \\
\hline
1   & square     & 0&.440684(1) & 1&.7496(1) &0&.440690(5)& 1&.0002(2) \\
    & triangular & 0&.27466(1)  & 1&.750(2)  & 0&.27466(1)& 1&.0005(5)  \\
    & honeycomb  & 0&.65849(2)  & 1&.750(1)  & 0&.65846(2)& 1&.000(1)  \\
    & Eq.~(\protect\ref{nm2expf})   & & & 1&.75      &     &      & 1&  \\
\hline
$\case{3}{2} $& square     &  0&.4804(2)  & 2&.30(1)   & 0&.4802(2) & 1&.31(2) \\
    & triangular &  0&.2967(1)  & 2&.30(2)   & 0&.2965(1) & 1&.31(1) \\
    & honeycomb  &  0&.73371(8) & 2&.313(5)  & 0&.7337(3) & 1&.33(1) \\
    & Eq.~(\protect\ref{nm2expf})   &  && 2&.31987...&    &       & 1&.33672... \\
\hline
$\case{7}{4} $& square     &  0&.5072(4)  & 2&.97(5)   & 0&.5066(4) & 1&.66(4) \\
    & triangular &  0&.3114(2)  & 2&.91(4)   & 0&.3111(5) & 1&.65(9) \\
    & honeycomb  &  0&.7844(5)  & 3&.01(5)   & 0&.7845(10)& 1&.74(9) \\
    & Eq.~(\protect\ref{nm2expf})   &  && 3&.12490...&   &        & 1&.80413... \\
\hline
$\case{19}{10}$& square    &  0&.529(1)   & 4&.0(2)    & *& & *&  \\
    & triangular &  0&.3230(4)  & 3&.7(1)    & 0&.323(2)  & 2&.2(3)  \\
    & honeycomb  &  0&.826(2)   & 4&.0(2)    & 0&.827(4)  & 2&.4(3)  \\
    & Eq.~(\protect\ref{nm2expf})   &  && 4&.72210...&  &         & 2&.72322... \\
\end{tabular}
\end{table}

\begin{table}
\squeezetable
\caption{
First-order integral-approximant analysis of the 20th order
strong-coupling series of $l_\chi\equiv \beta^{-1}\ln \chi$ on the
square lattice.  Asterisks mark defective approximants, i.e., those
having spurious singularities close to the real axis for
${\rm Re}\beta\protect\lesssim \beta_c$.
\label{lchisqIA}}
\begin{tabular}{rrrrr@{}lr@{}l}
\multicolumn{1}{r}{$N$}&
\multicolumn{1}{r}{$m$}&
\multicolumn{1}{r}{$l$}&
\multicolumn{1}{r}{$k$}& 
\multicolumn{2}{c}{$\beta_c$}&
\multicolumn{2}{c}{$\sigma$}\\
\tableline \hline
19 & 6 & 6 & 5 & 0&.5598 & 0&.55 \\
   & 6 & 5 & 6 &  &* & & \\
   & 7 & 5 & 5 & 0&.5563 & 0&.42 \\\hline
20 & 6 & 6 & 6 & 0&.5598 & 0&.59 \\
   & 7 & 6 & 5 & 0&.5585 & 0&.51 \\
   & 7 & 5 & 6 & 0&.5565 & 0&.37 \\
\end{tabular}
\end{table}

\begin{table}
\squeezetable
\caption{On the square lattice,
IA analysis of the series $\lambda_\chi$ constructed from the 
series of $l_\chi$ and $l_\chi^2$ according to  CPRM. 
$\sigma_{\rm biased}$
is obtained by biasing $x_c=1$. 
\label{blchisqIA}}
\begin{tabular}{rrrrr@{}lr@{}lr@{}l}
\multicolumn{1}{r}{$N$}&
\multicolumn{1}{r}{$m$}&
\multicolumn{1}{r}{$l$}& 
\multicolumn{1}{r}{$k$}& 
\multicolumn{2}{c}{$x_c$}&
\multicolumn{2}{c}{$\sigma$}&
\multicolumn{2}{c}{$\sigma_{\rm biased}$}\\
\tableline \hline
18 & 6 & 5 & 5 &  &*      &  &    & 0&.522 \\\hline
19 & 5 & 6 & 6 & 1&.0082 & 0&.64 & 0&.509 \\
   & 6 & 6 & 5 & 1&.0026 & 0&.55 & 0&.485 \\
   & 6 & 5 & 6 &  &*      &  &    & 0&.536 \\\hline
20 & 6 & 6 & 6 & 1&.0042 & 0&.57 &  &* \\
   & 6 & 7 & 5 & 1&.0043 & 0&.59 & 0&.506 \\
   & 6 & 5 & 7 & 1&.0198 & 1&.05 & 0&.501 \\
   & 7 & 6 & 5 & 1&.0039 & 0&.58 & 0&.523 \\
   & 7 & 5 & 6 & 1&.0035 & 0&.59 & 0&.467 \\
\end{tabular}
\end{table}

\begin{table}
\squeezetable
\caption{On the triangular lattice,
analysis of the series $\lambda_\chi$ constructed from the 
series of $l_\chi$ and $l_\chi^2$ according to  CPRM.
$\sigma_{\rm biased}$ is obtained by biasing $x_c=1$.
We noted here that sometime biasing the singularity at $x_c=1$
gives rise to spurious singularities in the real axis
for $x_c\protect\lesssim 1$. We considered approximants 
with singularities in the region [0.8,1.2] defective, and they are marked
by an asterisk. In these cases the
estimate of the exponent from non-biased approximants should be  more
reliable.
\label{blchitrIA}}
\begin{tabular}{rrrrr@{}lr@{}lr@{}l}
\multicolumn{1}{r}{$N$}&
\multicolumn{1}{r}{$m$}&
\multicolumn{1}{r}{$l$}& 
\multicolumn{1}{r}{$k$}& 
\multicolumn{2}{c}{$x_c$}&
\multicolumn{2}{c}{$\sigma$}&
\multicolumn{2}{c}{$\sigma_{\rm biased}$}\\
\tableline \hline
11 & 3 & 3 & 3 & 0&.9723 & 0&.192 & 0&.501 \\\hline
12 & 3 & 3 & 4 & 0&.9808 & 0&.258 & 0&.504 \\
   & 3 & 4 & 3 & 0&.9881 & 0&.335 & 0&.503 \\
   & 4 & 3 & 3 & 0&.9992 & 0&.490 & 0&.502 \\\hline
13 & 3 & 4 & 4 & 0&.9954 & 0&.446 & 0&.510 \\
   & 4 & 3 & 4 & 1&.0017 & 0&.534 & 0&.518 \\
   & 4 & 4 & 3 & 1&.0027 & 0&.549 &  &* \\
   & 5 & 3 & 3 & 1&.0026 & 0&.541 & 0&.493 \\\hline
14 & 4 & 4 & 4 & 1&.0016 & 0&.532 & 0&.550 \\
   & 4 & 5 & 3 & 1&.0003 & 0&.484 & 0&.477 \\
   & 4 & 3 & 5 & 1&.0015 & 0&.533 & 0&.485 \\
   & 5 & 3 & 4 & 1&.0015 & 0&.533 &  &* \\
   & 5 & 4 & 3 & 1&.0021 & 0&.520 & 0&.520 \\
   & 6 & 3 & 3 & 1&.0016 & 0&.523 & 0&.481 \\
\end{tabular}
\end{table}

\begin{table}
\squeezetable
\caption{
The strong-coupling estimates of $\xi_G$ are compared with some
available Monte Carlo results on the square lattice, taken from
Ref.~\protect\cite{Gupta} and obtained by standard Monte Carlo, and
Ref.~\protect\cite{Kim} by finite-size scaling techniques.  The
strong-coupling estimates of $\xi_G$, $\xi_G^{(1)}$ and $\xi_G^{(2)}$,
are obtained respectively from [9/9],[10/9],[9,10],[8/11] PA's of
$l_\xi^2$, and from [5/6/6], [6/6/5], [6/5/6], [5/6/5], [5/5/6],
[6/5/5] IA's of $l_\xi$ biased at $\beta_c = 0.559$.  We again warn
that the errors displayed in the strong-coupling estimates are related
to the spread of the different approximants considered.  The asterisk
indicates that the number concerns $\xi_{\rm exp}$, and not $\xi_G$.
\label{mc}}
\begin{tabular}{cr@{}lr@{}lr@{}lr@{}l}
\multicolumn{1}{c}{$\beta$}&
\multicolumn{2}{c}{$\xi_G^{_{(1)}}$}&
\multicolumn{2}{c}{$\xi_G^{_{(2)}}$}&
\multicolumn{2}{c}{$\xi_G^{_{({\rm MC})}}$\cite{Gupta}}&
\multicolumn{2}{c}{$\xi_G^{_{({\rm FSS})}}$\cite{Kim}}\\
\tableline \hline
1/2.2  &  9&.320(3)& 9&.318(1)  & 9&.32(2)$^*$  && \\    
1/2.08 & 18&.76(3) & 18&.74(1)  &18&.75(6) && \\ 
1/2.04 & 26&.3(1)  & 26&.21(2)  &26&.4(2)$^*$ && \\
0.5    & 40&.3(3)  & 40&.08(6)  &40&.4(4)$^*$  && \\
1/1.98 & 52&.2(3)  & 52&.0(1)  &51&.3(9)$^*$  && \\ 
1/1.96 & 70&.8(7)  & 70&.3(2)  &69&.9(8)  & 70&.4(4)      \\
1/1.94 & 102&(2)   & 100&.7(5) &&         &100&.3(7)      \\
1/1.92 & 158&(3)   & 156&(1)   & &         &156&(2)      \\
1/1.90 & 276&(11)  & 269&(3)   & &         &263&(3)      \\
1/1.88 & 570&(30)  & 549&(10)  & &         &539&(5)      \\
1/1.87 & 910&(60)  & 860&(20)  & &         &847&(7)      \\
\end{tabular}
\end{table}

\begin{table}
\squeezetable
\caption{
Summary of the determinations of $\beta_c$ and $\sigma$
 on the square, triangular and honeycomb lattices
by different analysis.
A bias in the analysis is indicated by a subscript
in the corresponding abbreviation. 
\label{summary}}
\begin{tabular}{cclr@{}lr@{}l}
\multicolumn{1}{c}{lattice}&
\multicolumn{1}{c}{series}&
\multicolumn{1}{c}{analysis}&
\multicolumn{2}{c}{$\beta_c$}&
\multicolumn{2}{c}{$\sigma$}\\
\tableline \hline
square 
&$l_\chi$      &  DLPA  &  0&.560(2)   &     0&.53(4) \\
&              &  IA    &  0&.558(2)   &     0&.49(8) \\
&$\lambda_\chi$&  CPRM-DLPA$_{x_c=1}$  &  && 0&.51(4) \\
&              &  CPRM-IA$_{x_c=1}$    &  && 0&.50(2) \\
&$\lambda_\xi$ &  CPRM-IA$_{x_c=1}$    &  && 0&.59(6) \\
&$l_\chi^2$    &  PA    &  0&.5579(3)  & & \\
&$l_\chi$      &  IA$_{\sigma=1/2}$    &  0&.5583(2)  & & \\
&$l_\xi^2$     &  PA    &  0&.558(1)   & & \\
&$l_\xi$       &  IA$_{\sigma=1/2}$    &  0&.559(1)  & & \\\hline
triangular
&$l_\chi$      &  DLPA  &  0&.3413(3)  &     0&.52(2) \\
&              &  IA    &  0&.33986(4) &     0&.473(3) \\
&$\lambda_\chi$&  CPRM-DLPA$_{x_c=1}$  &  && 0&.53(1) \\
&              &  CPRM-IA$_{x_c=1}$    &  && 0&.50(2) \\
&$\lambda_\xi$ &  CPRM-IA$_{x_c=1}$    &  && 0&.52(4) \\
&$l_\chi^2$    &  PA    &  0&.3400(2)  & & \\
&$l_\xi^2$     &  PA    &  0&.3393(1)  & & \\
&$l_\xi$       &  IA$_{\sigma=1/2}$    &  0&.3410(5)  & & \\\hline
honeycomb
&$l_\chi$      &  DLPA  &  0&.884(1)  & 0&.54(1) \\
&                     &  IA    &  0&.877(6)  & 0&.4(2) \\
&$l_\chi^2$    &  PA    &  0&.879(1)  & & \\
&$l_\chi  $    &  IA$_{\sigma=1/2}$    &  0&.880(1)  & & \\
&$l_\xi^2$     &  PA    &  0&.878(2)  & & \\
&$l_\xi$       &  IA$_{\sigma=1/2}$    &  0&.883(1)  & & \\
\end{tabular}
\end{table}

\begin{table}
\squeezetable
\caption{
We report $s^*$, $c_2$ and $c_3$ as obtained from the analysis of the
strong-coupling series in $\beta$ (first line) and in $E$ (second
line) on the square, triangular and honeycomb lattices.  Beside the
spread of estimates from different PA's and Dlog-PA's, the errors
displayed take also into account the uncertainty on $\beta_c$ and
$E_c$.  We do not report the estimates of $c_3$ by the analysis of the
$\beta$-series because their uncertainty is much larger than those
from the $E$-series.
\label{tabsuN2}}
\begin{tabular}{clr@{}lr@{}lr@{}l}
\multicolumn{1}{c}{lattice}&
\multicolumn{1}{c}{$\beta_c,E_c$}&
\multicolumn{2}{c}{$s^*$}&
\multicolumn{2}{c}{$c_2$}&
\multicolumn{2}{c}{$c_3$}\\
\tableline \hline
square    & $\beta_c\simeq 0.559$ &  0&.9985(12)  & 0&.000(2) & &\\
          & $E_c\simeq 0.722$     &  0&.9984(7)   &-0&.0014(3)& 0&.00001(2) \\\hline
triangular& $\beta_c\simeq 0.340$ &  0&.9979(11)  &-0&.002(2) & &\\
          & $E_c\simeq 0.68$      &  0&.9985(11)  &-0&.0010(3)& 0&.00001(5)\\\hline
honeycomb & $\beta_c\simeq 0.880$ &  0&.9988(10)  &-0&.001(3) & &\\
          & $E_c\simeq 0.77$      &  0&.9987(5)   &-0&.0021(4)& 0&.00003(2) \\
\end{tabular}
\end{table}

\begin{table}
\squeezetable
\caption{
For $N=1$ and $N=0$ we report $s^*$, $c_2$ and $c_3$ as obtained from
the analysis of the available strong-coupling series in $\beta$ on the
square, triangular and honeycomb lattices.  The analyses of the
corresponding energy series provide consistent but less precise
results, so we do not report their results.
\label{tabsuN10}}
\begin{tabular}{rccr@{}lr@{}lr@{}l}
\multicolumn{1}{c}{$$}&
\multicolumn{1}{c}{lattice}&
\multicolumn{1}{c}{$\beta_c$}&
\multicolumn{2}{c}{$s^*$}&
\multicolumn{2}{c}{$c_2$}&
\multicolumn{2}{c}{$c_3$}\\
\tableline \hline
N=1 &  square     & ${1\over 2}\ln(\sqrt{2}+1)=0.440687...$  &  
                    0&.99909(2)  &  -0&.00094(3) & 0&.000008(5) \\
    &  triangular & ${1\over 4}\ln 3=0.274653...$            &  
                    0&.99912(5) &  -0&.00098(4) & 0&.00001(1)  \\ 
    &  honeycomb  & ${1\over 2}\ln(2+\sqrt{3})=0.658478...$  &  
                    0&.99907(2)  &  -0&.00093(3) & 0&.000012(2) \\ \hline
N=0 &  square     & 0.3790527(2)\cite{GuttmannO0}       &  
                    1&.0001(2)   &  0&.00016(8) & 0&.00000(1) \\
    &  triangular & 0.240920(1)\cite{GuttmannO0}        &  
                    1&.0002(4)   &  0&.0003(5) &  0&.00000(3) \\
    &  honeycomb  & $(2+\sqrt{2})^{-1/2}=0.541196...$\cite{Nienhuis}&  
0&.9998(2)   &  0&.00010(7) & -0&.00002(1)\\ 
\end{tabular}
\end{table}

\end{document}